\documentclass[onecolumn, draftclsnofoot, journal]{IEEEtran}
\usepackage[cmex10]{amsmath}
\usepackage{amssymb}
\usepackage{amsthm}
\usepackage{color}
\usepackage[dvipdfmx]{graphicx}
\usepackage[abs]{overpic}
\usepackage{subcaption}
\usepackage{array}
\usepackage{cite}
\usepackage{url}

\usepackage{amsmath}
\usepackage{amssymb}
\usepackage{amsthm}
\theoremstyle{definition}
\newtheorem{thm}{Theorem}
\newtheorem{lem}{Lemma}
\newtheorem{cor}{Corollary}
\newtheorem{defn}{Definition}
\newtheorem{rem}{Remark}
\newcommand{\ra}{\rightarrow}

\newcommand{\teq}{\triangleq}
\newcommand{\A}{\alpha}

\newcommand{\vep}{\varepsilon}

\newcommand{\cD}{\mathcal{D}}
\newcommand{\cE}{\mathcal{E}}

\newcommand{\cP}{\mathcal{P}}

\newcommand{\cR}{\mathcal{R}}
\newcommand{\cS}{\mathcal{S}}

\newcommand{\cX}{\mathcal{X}}
\newcommand{\cY}{\mathcal{Y}}

\newcommand{\bfd}{\mathbf{d}}

\newcommand{\bfX}{\mathbf{X}}
\newcommand{\bfY}{\mathbf{Y}}

\newcommand{\barX}{\bar{X}}
\newcommand{\barY}{\bar{Y}}

\def\eqo#1{\overset{\mathrm{#1}}{=}}
\def\geqo#1{\overset{\mathrm{#1}}{\geq}}
\def\leqo#1{\overset{\mathrm{#1}}{\leq}}

\allowdisplaybreaks

\ifCLASSINFOpdf
\else
\fi
\begin{document}
%
\title{Coding Theorems for Asynchronous Slepian-Wolf Coding Systems\thanks{This work was supported in part by JSPS KAKENHI Grant Numbers JP15K15935 and JP19K04368. This paper was presented in part at a Technical Meeting of Technical Committee on Information Theory \cite{matsuta2013universal} and the 2015 IEEE Information Theory Workshop \cite{matsuta2015achievable}.}}
%
%
%

\author{Tetsunao~Matsuta,~\IEEEmembership{Member,~IEEE,}
  and~Tomohiko~Uyematsu,~\IEEEmembership{Senior~Member,~IEEE}
  \thanks{T. Matsuta and T. Uyematsu are with the Department of Information and Communications Engineering, Tokyo Institute of Technology, Ookayama, Meguro-ku, Tokyo, 152-8552, Japan (e-mail: matsuta@ieee.org, uyematsu@ieee.org).}
}

\maketitle

\begin{abstract}
    The Slepian-Wolf (SW) coding system is a source coding system with two encoders and a decoder, where these encoders independently encode source sequences from two correlated sources into codewords, and the decoder reconstructs both source sequences from the codewords. In this paper, we consider the situation in which the SW coding system is asynchronous, i.e., each encoder samples a source sequence with some unknown delay. We assume that delays are unknown but maximum and minimum values of possible delays are known to encoders and the decoder. We also assume that sources are discrete stationary memoryless and the probability mass function (PMF) of the sources is unknown but the system knows that it belongs to a certain set of PMFs. For this asynchronous SW coding system, we clarify the achievable rate region which is the set of rate pairs of encoders such that the decoding error probability vanishes as the blocklength tends to infinity. We show that this region does not always coincide with that of the synchronous SW coding system in which each encoder samples a source sequence without any delay.
\end{abstract}

\begin{IEEEkeywords}
    Achievable rate region, asynchronous, Slepian-Wolf coding, universal coding scheme.
\end{IEEEkeywords}

%
\IEEEpeerreviewmaketitle

\section{Introduction}
\label{sec:introduction}
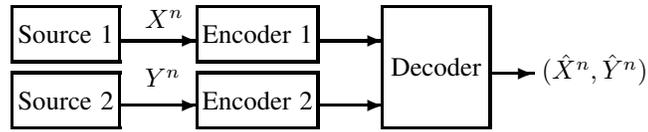
\begin{figure}[!t]
    \centering
    \begin{picture}(230, 40)
        \thicklines

        \put(40, 32.5){\vector(1, 0){30}}
        \put(50, 37.5){$X^n$}
        \put(40, 8){\vector(1, 0){30}}
        \put(50, 15){$Y^n$}
        \put(115, 32.5){\vector(1, 0){25}}
        \put(115, 8){\vector(1, 0){25}}
        \put(180, 20){\vector(1, 0){17}}
        \put(200, 18){$(\hat X^n,\hat Y^n)$}

        \put(0,25){\framebox(40, 20){Source 1}}
        \put(0,0){\framebox(40, 20){Source 2}}
        \put(70,25){\framebox(45, 20){Encoder 1}}
        \put(70,0){\framebox(45, 20){Encoder 2}}
        \put(140,0){\framebox(40, 45){Decoder}} 
    \end{picture}
    \caption{Slepian-Wolf coding system}
    \label{fig: SWF}
\end{figure}

\IEEEPARstart{T}{he} Slepian-Wolf (SW) coding system \cite{slepian1973ncc} is one of famous source coding systems with many terminals. In this coding system (see Fig.\ \ref{fig: SWF}), two encoders independently encode source sequences from two correlated sources into codewords, and the decoder reconstructs both source sequences from the codewords. For this coding system, Slepian and Wolf \cite{slepian1973ncc} characterized the achievable rate region for discrete stationary memoryless sources (DMSs), where the achievable rate region is the set of rate pairs of encoders such that the decoding error probability vanishes as the blocklength tends to infinity.

Discrete-time source symbols are regarded as discrete-time samples of a discrete-time process such as coin flips or a continuous-time process such as a wave. In the above SW coding system, it is assumed that these processes are sampled at the encoders without delay. Thus, encoders can encode a pair of source sequences with an expected correlation. In other words, two encoders are \textit{synchronous}. However, in practice, the encoders are not always synchronous. It is natural to assume that these processes are sampled at encoders with some unknown delays. In other words, the encoders are \textit{asynchronous}. There are two reasons (i) and (ii) to justify this assumption: (i) Since the encoders are independent in the system and cannot exchange any information, it is difficult to completely adjust the time to start sampling among the encoders. Thus, in general, uncontrollable unknown delays occur. (ii) Even if the encoders can adjust the time to start sampling by exchanging information or referring a shared clock, the encoders become asynchronous when processes arrive late to the encoders. An example of this case is as follows: Observatories (encoders) on islands sample wave heights (source sequences) per unit time caused by breeze, an earthquake, a typhoon, etc. Since islands are separated, a wave reaches an island later than it reaches the other island. The observatories send the sequences to a weather center (decoder) on a coast city distant from there. In this example, the observatories (and also the weather center) do not know the actual delay of the wave in advance because there are many uncertainties such as the direction of breeze, the point of the earthquake center, shielding on the sea, etc. Thus, even if observatories can adjust the time to start sampling, they sample the wave with some unknown delay.

\begin{figure*}[!t]
    \begin{subfigure}{1\textwidth}
        \centering
        \begin{overpic}[height=40mm]{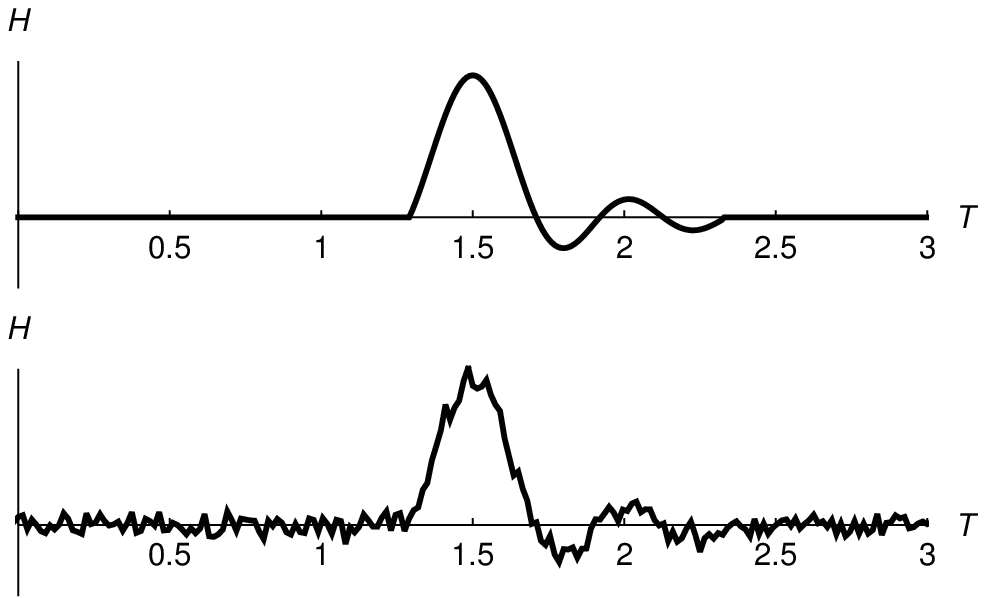}
            \multiput(58,110)(0,-2){56}{\line(0,-1){1}}
            \multiput(115.6,110)(0,-2){56}{\line(0,-1){1}}
            \multiput(173.2,110)(0,-2){56}{\line(0,-1){1}}
        \end{overpic}
        \caption{Waves without a delay}
        \label{subfig: wave 1}
    \end{subfigure}
    \begin{subfigure}{.49\textwidth}        
        \begin{overpic}[height=40mm]{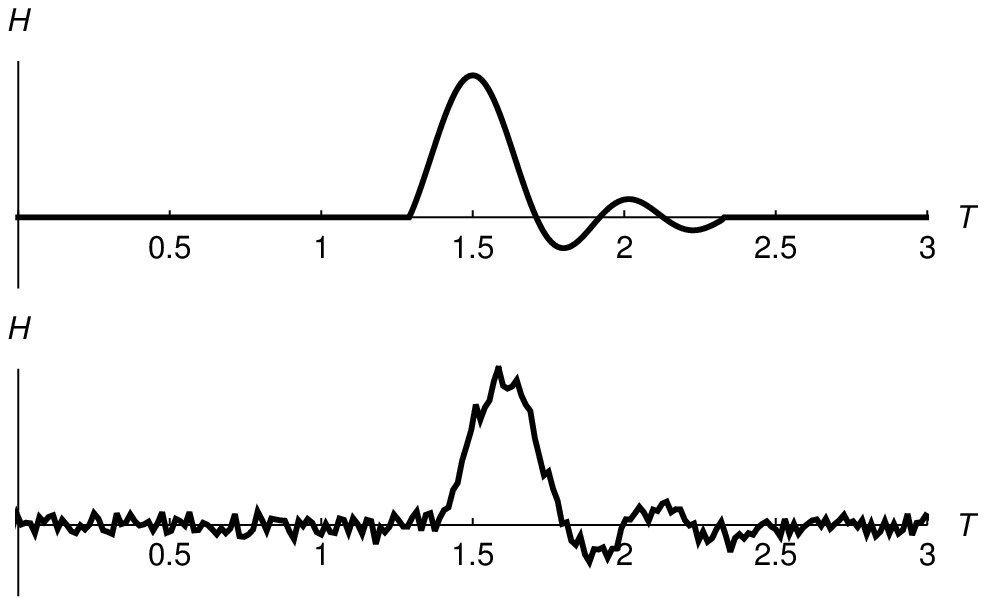}
            \multiput(58,110)(0,-2){56}{\line(0,-1){1}}
            \multiput(115.6,110)(0,-2){56}{\line(0,-1){1}}
            \multiput(173.2,110)(0,-2){56}{\line(0,-1){1}}
        \end{overpic}
        \caption{Waves with a tiny delay}
        \label{subfig: wave 2}
    \end{subfigure}
    \begin{subfigure}{.49\textwidth}
        \begin{overpic}[height=40mm]{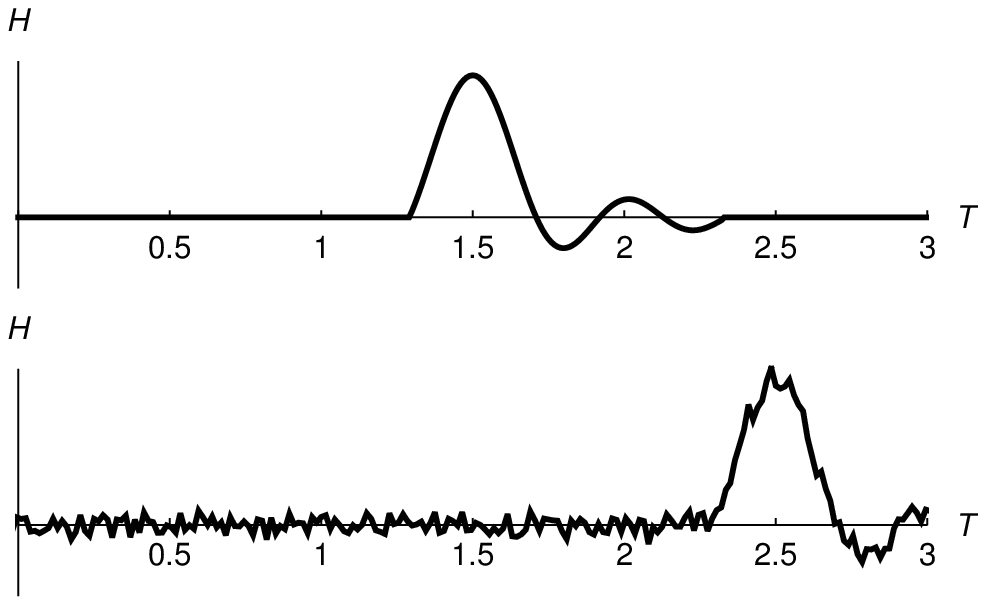}
            \multiput(58,110)(0,-2){56}{\line(0,-1){1}}
            \multiput(115.6,110)(0,-2){56}{\line(0,-1){1}}
            \multiput(173.2,110)(0,-2){56}{\line(0,-1){1}}
        \end{overpic}
        \caption{Waves with a unit delay}
        \label{subfig: wave 3}
    \end{subfigure}
    \caption{Asynchronicity of sampling}
    \label{fig: waves}
\end{figure*}

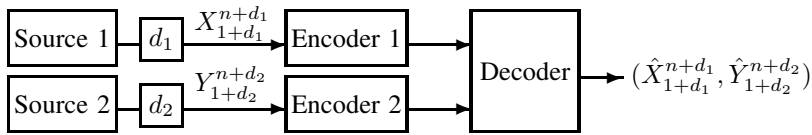
\begin{figure*}[!t]
    \centering
    \begin{picture}(300, 40)
        \thicklines

        \put(40, 32.5){\line(1, 0){10}}
        \put(65, 32.5){\vector(1, 0){40}}
        \put(70, 38){$X_{1+d_{1}}^{n+d_{1}}$}
        \put(40, 8){\line(1, 0){10}}
        \put(65, 8){\vector(1, 0){40}}
        \put(70, 15){$Y_{1+d_{2}}^{n+d_{2}}$}
        \put(150, 32.5){\vector(1, 0){25}}
        \put(150, 8){\vector(1, 0){25}}
        \put(215, 20){\vector(1, 0){17}}
        \put(235, 18){$(\hat X_{1+d_{1}}^{n+d_{1}},\hat Y_{1+d_{2}}^{n+d_{2}})$}
        
        \put(0,25){\framebox(40, 20){Source 1}}
        \put(50,27){\framebox(15, 15){$d_1$}}
        \put(0,0){\framebox(40, 20){Source 2}}
        \put(50,2){\framebox(15, 15){$d_2$}}
        \put(105,25){\framebox(45, 20){Encoder 1}}
        \put(105,0){\framebox(45, 20){Encoder 2}}
        \put(175,0){\framebox(40, 45){Decoder}} 
    \end{picture}
    \caption{Slepian-Wolf coding system with delays $d_1$ and $d_2$, where $d_1, d_2 \in \{\cdots, -2, -1, 0, 1, 2 \cdots\}$}
    \label{fig: SWFwithDs}
\end{figure*}

To make matters worse, unknown delays may cause uncertainty of statistical properties of sources. To justify this, we give an example where discrete-time source symbols are quantized and sampled version of a continuous signal. Let us consider correlated wave signals shown in Fig.\ \ref{fig: waves} (\subref{subfig: wave 1}). The upper wave is merely a constant wave in which sometimes large changes occur. The lower wave is its noisy version where a Gaussian process is added. In this example, we assume that each wave is sampled at each encoder per unit time at dotted lines and quantized with some resolution. We also assume that a large change does not affect two unit times. These sampled and quantized signals can be regarded as discrete-time and discrete-valued source symbols. If waves are sampled without any delay (see Fig.\ \ref{fig: waves} (\subref{subfig: wave 1})), the encoders are synchronous. In that case, the correlation between two symbols is characterized by only one channel induced by a Gaussian process. On the other hand, if waves  are sampled with a tiny delay as shown in Fig.\ \ref{fig: waves} (\subref{subfig: wave 2}), the correlation between two symbols no longer corresponds to that of the case without delay. When the delay is unknown, this causes uncertainty of statistical properties of sources, and now the properties are characterized by a set of channels rather than a singleton i.e., only one channel. We also note that when the delay corresponds to a unit time as shown in Fig.\ \ref{fig: waves} (\subref{subfig: wave 3}), the encoders encode source sequences with integer-valued delays as shown in Fig.\ \ref{fig: SWFwithDs}.

As a consequence of the discussion so far, the asynchronous SW coding system can be represented as the SW coding system with integer-valued delays (Fig.\ \ref{fig: SWFwithDs}), where the delays and statistical properties of sources are unknown. Thus, in what follows, a delay refers to an integer-valued delay unless otherwise stated. We note that when a continuous signal is not assumed behind source symbols, the case where delays are unknown but statistical properties are known is also worth considering. We also note that, in general, uncertainty of statistical properties does not only come from unknown (real-valued) delays.

There are some related studies to the asynchronous SW coding system in the case where statistical properties of sources are \textit{known} in advance. Willems \cite{willems1988totally} considered the situation in which delays are unknown to encoders but known to the decoder. He showed that the achievable rate region for DMSs coincides with that of the synchronous SW coding system. In the same assumption as \cite{willems1988totally}, Rimoldi and Urbanke \cite{rimoldi1997asynchronous} and Sun et al.\ \cite{5407617} gave coding schemes based on source splitting. In these studies \cite{willems1988totally,rimoldi1997asynchronous,5407617}, it is implicitly assumed that for a given finite blocklength, the encoders continue to transmit codewords infinitely and the decoder has infinitely large memory to receive those infinitely many codewords. Most importantly, this assumption eliminates the effect of delays because the decoder can wait infinitely long time until receiving correlated codewords even if the blocklength is finite. However, in practice, the decoder cannot have infinitely large memory and wait for decoding infinitely long time. Moreover, when delays are very large, the decoding delay is also very large regardless of the blocklength. This justifies considering the coding system encoding only one pair of source sequences for a given blocklength as the system shown in Fig.\ \ref{fig: SWFwithDs}. In this coding system, the decoder outputs an estimation from a pair of codewords and does not wait for the next pair. Obviously, this system includes the case where a codeword consists of sub-codewords. Thus, the decoder waits until it receives a pair of codewords or finite pairs of sub-codewords, and does not wait infinitely long time for a given finite blocklength. Here, for fairness among encoders and simplicity of the system, the blocklength of source sequences are assumed to be the same. Oki and Oohama \cite{oki1997coding} considered this coding system, i.e., the system shown in Fig.\ \ref{fig: SWFwithDs}, where statistical properties of sources are known in advance. They assumed that delays are unknown to encoders and also the decoder, but maximum and minimum values of possible delays are bounded and known to the decoder. They showed that the achievable rate region for DMSs coincides with that of the synchronous SW coding system.

In this paper, we also consider the asynchronous SW coding system. Specifically, we consider the SW coding system with delays under the following two assumptions (i) and (ii): (i) Delays are unknown but maximum and minimum values of possible delays are known to encoders and the decoder. In the above example of islands, the maximum delay depends on the distance between the islands and is naturally known to the observatories and the weather center because the distance is known in advance. (ii) Sources are DMSs and the probability mass function (PMF) of the sources is unknown but a set of PMFs including it is known to encoders and the decoder. Unlike the assumption in \cite{oki1997coding}, we allow delays to be unbounded and maximum and minimum values of possible delays to be subject to change by the blocklength. This allows us more detailed analyses such as the case where delays affect a half of source sequences, the case where a delay always occurs, etc. This can also be seen as the following situation: Each encoder has a FIFO (or LIFO) memory (i.e., a source sequence is always new or old). Since the encoding and decoding delay of a preceding sequence has an order depending on the blocklength, which sequence is stored in the memory depends on the blocklength. Consequently, possible delays also depend on the blocklength.

For this asynchronous SW coding system, we clarify the achievable rate region and show that the region does not always coincide with that of the synchronous SW coding system. This result is completely different from results of the above related studies. We use a usual information-theoretic technique as in \cite{cover2006eit} to the proof of the converse part. On the other hand, the direct part is the challenging part, and its proof is somewhat different from the usual one. Instead of directly dealing with coding for sources with delays, we deal with coding for a \textit{mixed} source given by the mixture of all sources with possible delays and employ Gallager's random coding techniques \cite{gallager1968information} and \cite{gallager1976} to the mixed source. We note that given encoders and a decoder are \textit{universal} in the sense that they can encode and decode (asymptotically) correctly even if statistical parameters such as the delays and the PMF of DMSs are unknown. We used an analogous technique using a mixed source in \cite{matsuta2010universalIEICE} to prove the existence of a code. However, since we could not use Gallager's techniques to the mixed source in \cite{matsuta2010universalIEICE}, we did not give an exponential bound in it. We also give an extension of our coding scheme: If possible delays are bounded, our coding scheme can be extended to a scheme which does not require knowledge of the actual bound of delays. We note that this is also an extension of the result by Oki and Oohama \cite{oki1997coding}.

The rest of this paper is organized as follows. In Section \ref{sec:preliminaries}, we give some notations and the formal definition of the asynchronous SW coding system and the achievable rate region. In Section \ref{sec:AchievableRateRegion}, we show the achievable rate region and some properties of it. In Sections \ref{sec:converse part} and \ref{sec:direct part}, we show the converse part and the direct part to clarify the achievable rate region, respectively. In Section \ref{sec:elimination_knowledge}, we give the extension of our coding scheme. In Section \ref{sec:conclusion}, we conclude the paper.

\section{Preliminaries}
\label{sec:preliminaries}
In this section, we provide some notations and the precise definition of the asynchronous SW coding system.

We will denote an $n$-length sequence of symbols $(a_1,a_2,\cdots,a_n)$ by $a^n$, a sequence of symbols $(a_{m},a_{m+1},\cdots,a_{m'})$ by $a_m^{m'}$, and a pair of sequences of symbols $((a_{m},b_{l})$, $(a_{m+1},b_{l+1})$, $\cdots$, $(a_{m'}, b_{l'}))$ by $(a_m^{m'}, b_{l}^{l'})$. If $m > m'$, we assume $a_m^{m'} = \emptyset$. For any finite sets $\cX$ and $\cY$, we will denote the set of all PMFs over $\cX$ by $\cP(\cX)$, and the set of all conditional PMFs over $\cY$ given elements of $\cX$ by $\cP(\cY|\cX)$. Unless otherwise stated, the PMF of a random variable (RV) $X$ on $\cX$ 
will be denoted by $P_{X}\in\cP(\cX)$,
and the conditional PMF of $Y$ on $\cY$ given $X$
will be denoted by $P_{Y|X}\in\cP(\cY|\cX)$.
We will denote the $n$th power of a PMF $P_X$ by $P_X^n$, i.e.,
$P_X^n(x^n)=\prod_{i=1}^n P_{X}(x_i)$,
and the $n$th power of a conditional PMF $P_{Y|X}$ by $P_{Y|X}^n$, i.e.,
$P_{Y|X}^n(y^n|x^n)=\prod_{i=1}^n P_{Y|X}(y_i|x_i)$.  In what follows, all logarithms and exponentials are taken to the base 2.

We assume that $\cX$ and $\cY$ are finite sets. We will denote a general source $\{(X^{n},Y^{n})\}_{n=1}^\infty$ (i.e., a sequence of $n$-length RVs) by the corresponding boldface letter $(\bfX, \bfY)$ (cf.\ \cite{hanspringerinformation}). Since a pair of DMSs is represented by a sequence of independent copies of a pair of RVs $(X,Y)$, we simply write it as $(X,Y)$.

In the asynchronous SW coding system, two $n$-length sequences from DMSs $(X,Y)$ are independently encoded by encoder 1 and encoder 2, respectively. Hence, for positive integers $M_n^{(1)}$ and $M_n^{(2)}$, encoder 1 and encoder 2 are defined by the mappings
\begin{align*}
    & f_n^{(1)}:\mathcal{X}^n \to \mathcal{M}_n^{(1)}=\{1,\cdots,M_n^{(1)}\},\\
    & f_n^{(2)}:\mathcal{Y}^n \to \mathcal{M}_n^{(2)}=\{1,\cdots,M_n^{(2)}\},
\end{align*}
and the rates of these encoders are defined as
\begin{align*}
    R_n^{(1)} &\teq \frac{1}{n}\log{M_n^{(1)}},\\
    R_n^{(2)} &\teq \frac{1}{n}\log{M_n^{(2)}},
\end{align*}
respectively. Since the encoders are asynchronous, encoder 1 might encode a source sequence $X^n=(X_1,X_2,\cdots,X_{n})$ while encoder 2 might encode a source sequence $Y_{-2}^{n-3}=(Y_{-2},Y_{-1},\cdots,Y_{n-3})$. In general, encoder 1 and encoder 2 encode sequences
$X^n$ and $Y_{1+d}^{n+d}=(Y_{1+d},Y_{2+d},\cdots,Y_{n+d})$, respectively, where $d$ is an integer which represents a \textit{relative}\footnote[2]{Suppose that encoder 1 and encoder 2 run with delays $d_1$ and $d_2$, respectively. Then, for a source sequence $(X_{1+d_1}^{n+d_1},Y_{1+d_2}^{n+d_2})$ encoded by encoders, it holds that $P_{X_{1+d_1}^{n+d_1}Y_{1+d_2}^{n+d_2}}(x^n,y^n) = P_{X_{1}^{n}Y_{1+d_2-d_1}^{n+d_2-d_1}}(x^n,y^n)$. Hence, we may only consider a relative delay.} delay (see Tables \ref{table: correlation d geq 0} and \ref{table: correlation d leq 0}). 

\begin{table}[t]
    \caption{Correlation in the case where $0 \leq d$}
    \label{table: correlation d geq 0}
    \begin{center}
        \begin{tabular}{c|c|c|c|c|c|c|c}
          $X_1$ & $\cdots$ & $X_{1+d}$
          & $\cdots$ & $X_n$ & \multicolumn{3}{c}{$\cdots$} \\
          \hline
          \multicolumn{2}{c|}{$\cdots$} &$Y_{1+d}$ & $\cdots$
          & $Y_n$ & $Y_{n+1}$ & $\cdots$ & $Y_{n+d}$
        \end{tabular} 
    \end{center}

    \caption{Correlation in the case where $d \leq 0$}
    \label{table: correlation d leq 0}
    \begin{center}
        \begin{tabular}{c|c|c|c|c|c|c|c}
          \multicolumn{2}{c|}{$\cdots$} & $X_1$ & $\cdots$ & $X_{n+d}$
          & $X_{1+n+d}$ & $\cdots$ & $X_n$ \\
          \hline
          $Y_{1+d}$ & $\cdots$ & $Y_1$
                                                           & $\cdots$ &$Y_{n+d}$ & \multicolumn{3}{c}{$\cdots$}
        \end{tabular}
    \end{center}
\end{table}

Without loss of generality, we assume that $-n\leq d \leq n$ because, for any $d \geq n$ or $d \leq -n$, $X^n$ is independent of $Y_{1+d}^{n+d}$. We denote $Y_{1+d}^{n+d}$ by $Y_{(d)}^{n}$ for the sake of brevity. Note that
\begin{align}
    P_{X^{n}Y_{(d)}^{n}}(x^n,y^n)
    = P_{X}^{|d|}(x_{\rm I}^{|d|}) P_{XY}^{n-|d|}(x_{\rm C}^{n - |d|}, y_{\rm C}^{n-|d|}) P_{Y}^{|d|}(y_{\rm I}^{|d|}),
    \label{eq:dist. of a source with relative delays}
\end{align}
where $P_{XY}$ is the PMF of the pair of DMSs $(X,Y)$,
$P_{X}$ and $P_{Y}$ are the marginal PMFs of $P_{XY}$, and
\begin{align*}
    x_{\rm I}^{|d|}
    & =
    \begin{cases}
        x_1^{d} & \mbox{ if }0 \leq d\leq n ,\\
        x_{1+n+d}^n & \mbox{ if } - n\leq d\leq 0,
    \end{cases}\\
    (x_{\rm C}^{n-|d|},y_{\rm C}^{n-|d|})
    & =
    \begin{cases}
        (x_{1+d}^{n},y_{1}^{n-d}) & \mbox{ if }0 \leq d\leq n ,\\
        (x_{1}^{n+d},y_{1-d}^{n}) & \mbox{ if }-n\leq d\leq 0,
    \end{cases}\\
    y_{\rm I}^{|d|}
    & =
    \begin{cases}
        y_{1+n-d}^n & \mbox{ if }0 \leq d\leq n, \\
        y_1^{-d} & \mbox{ if }-n\leq d\leq 0.
    \end{cases}
\end{align*}
We denote DMSs $(X,Y)$ with a delay $d$ simply by $(X, Y_{(d)})$ which implies the sequence of RVs $\{(X^n,Y_{(d)}^{n})\}_{n=1}^\infty$. We introduce the maximum $\overline d_{n}$ $(\leq n)$ and the minimum $\underline d_{n}$ $(\geq -n)$ of possible delays, and denote the sequence $\{(\underline d_{n}, \overline d_{n})\}_{n=1}^\infty$ by $\bfd$. Hence, any possible delay $d$ satisfies $\underline d_{n}\leq d \leq \overline d_{n}$ for any blocklength $n$. We allow the maximum and the minimum of delays to be changed with the blocklength.

The decoder receives two codewords $f_n^{(1)}(X^{n})$ and $f_n^{(2)}(Y_{(d)}^{n})$,
and outputs an estimate of the pair of sequences $(X^n,Y_{(d)}^{n})$.
Hence, the decoder is defined by the mapping
\begin{align*}
    \varphi_n:\mathcal{M}_n^{(1)} \times \mathcal{M}_n^{(2)} \to \mathcal{X}^n\times\cY^n.
\end{align*}
Then, for DMSs $(X,Y)$ and a delay $d$,
the error probability is defined as
\begin{align}
    \vep_{XY_{(d)}}^{(n)}(f_n^{(1)}, f_n^{(2)}, \varphi_n) \teq \Pr\{\varphi_n(f_n^{(1)}(X^{n}), f_n^{(2)}(Y_{(d)}^{n})) \neq (X^{n}, Y_{(d)}^{n})\}. \label{eq:def er pro}
\end{align}
More generally, we will denote the error probability for a general source $(\bfX, \bfY)$ by $\vep_{\bfX\bfY}^{(n)}(f_n^{(1)}, f_n^{(2)}, \varphi_n)$, i.e.,
\begin{align*}
    \vep_{\bfX\bfY}^{(n)}(f_n^{(1)}, f_n^{(2)}, \varphi_n)
    \teq \Pr\{\varphi_n(f_n^{(1)}(X^{n}), f_n^{(2)}(Y^{n})) \neq (X^{n}, Y^{n})\}.
\end{align*}
We will sometimes omit $(f_n^{(1)}, f_n^{(2)}, \varphi_n)$ in the notation of $\vep_{\bfX\bfY}^{(n)}$ when it is clear from the context.

In this coding system, we assume that the actual delay $d$ is unknown but the bound $\bfd$ of delays is known to the encoders and the decoder. Furthermore, we assume that the PMF $P_{XY}$ of the pair of DMSs is an element of a given set $\cS \subseteq \cP(\cX \times \cY)$ of PMFs and that the PMF $P_{XY}$ is unknown but the set $\cS$ of PMFs is known to the encoders and the decoder. More precisely, the code $(f_n^{(1)}, f_n^{(2)}, \varphi_n)$ is independent of $d$ and $P_{XY}\in\cS$, but is allowed to be dependent on $\bfd$ and $\cS$. Hence, as we mentioned earlier, the code is universal in the sense that we require that they can encode and decode (asymptotically) correctly even if statistical parameters $d$ and $P_{XY}$ are unknown.

We now define \textit{achievability} and \textit{achievable rate region} for the asynchronous SW coding system for a given set $\cS$ of PMFs of sources and a bound $\bfd$ of delays.
\begin{defn}[Achievability]
    A pair $(R_1,R_2)$ is called \textit{achievable} for a set $\cS$ of PMFs and a bound $\bfd$ of delays if and only if there exists a sequence of codes $\{(f_n^{(1)}, f_n^{(2)}, \varphi_n)\}$ satisfying
    \begin{align*}
        \limsup_{n\ra\infty } R_n^{(1)} &\leq R_1,\\
        \limsup_{n\ra\infty } R_n^{(2)} &\leq R_2,
    \end{align*}
    and
    \begin{align}
        \lim_{n\ra\infty} \sup_{P_{XY} \in\cS} \max_{d \in\cD_{n}}
        \vep_{XY_{(d)}}^{(n)}(f_n^{(1)}, f_n^{(2)}, \varphi_n) = 0,
        \label{eq:def of achiv of error prob}
    \end{align}
    where $\cD_{n}\teq \{\underline d_{n}, \underline d_{n}+1,\cdots,\overline d_{n}-1,\overline d_{n}\}$ is the set of possible delays.
\end{defn}
\begin{defn}[Achievable rate region]
    For a set $\cS$ of PMFs and a bound $\bfd$ of delays,
    the achievable rate region $\cR_{\bfd}(\cS)$ is defined by
    \begin{align*}
        \cR_{\bfd}(\cS)\teq\{& (R_1,R_2):(R_1,R_2)\mbox{ is achievable for the set } \cS \mbox{ and the bound }\bfd\}.
    \end{align*}
    When $\cS = \{P_{XY}\}$ is a singleton, we simply write $\cR_{\bfd}(\cS)$ as $\cR_{\bfd}(X,Y)$. Moreover, when there is no delay, i.e, $\overline d_{n} = \underline d_{n} = 0$, we simply write $\cR_{\bfd}(X,Y)$ as $\cR(X,Y)$.
\end{defn}
\begin{rem}
    When $\cS$ is a singleton and $\overline{d}_{n} = \underline{d}_{n} = 0$, our coding system corresponds to the usual synchronous SW coding system. Hence, $\cR(X,Y)$ denotes the achievable rate region of the synchronous SW coding system.
\end{rem}
Oki and Oohama \cite{oki1997coding} considered the case where $\cS$ is a singleton and delays are bounded in the sense that for a constant $c\in[0,\infty)$, $-c = \underline d_{n}\leq d \leq \overline d_{n}= c,$ $\forall n > 0$. They assume that the delay is unknown to encoders and the decoder but $c$ is known to the decoder. For this special case, they showed that the achievable rate region coincides with that of the synchronous SW coding system, i.e.,
\begin{align}
    \cR_{\bfd}(X,Y) &= \cR(X,Y) \notag\\
    &= \{(R_1,R_2): R_1 \geq H(X|Y), R_2 \geq H(Y|X), R_1 + R_2 \geq H(X,Y) \},
    \label{equ: achievable rate region for synchronous SW coding systems}
\end{align}
where $H(X)$ denotes the entropy of $X$, and $H(X|Y)$ denotes the conditional entropy of $X$ given $Y$. We will also denote $H(X)$ by $H(P_{X})$, and $H(X|Y)$ by $H(P_{X|Y}|P_{Y})$ to clear their PMFs.

In what follows, we sometimes use the following notations for the sake of simplicity: $H_{1}(X, Y) = H(X|Y)$, $H_{2}(X, Y) = H(Y|X)$, $H_{3}(X, Y) = H(X, Y)$, and $R_3 = R_1 + R_2$. Due to these notations, \eqref{equ: achievable rate region for synchronous SW coding systems} is also written as
\begin{align*}
    \{(R_1,R_2): R_i \geq H_{i}(X, Y),\ \forall i \in \{1, 2, 3\}\}.
\end{align*}

\section{Achievable Rate Region}
\label{sec:AchievableRateRegion}
In this section, we show the achievable rate region of the asynchronous SW coding system. We also show that the obtained region does not always coincide with that of the synchronous SW coding system.

\begin{figure*}[!t]
    \centering
    \begin{overpic}{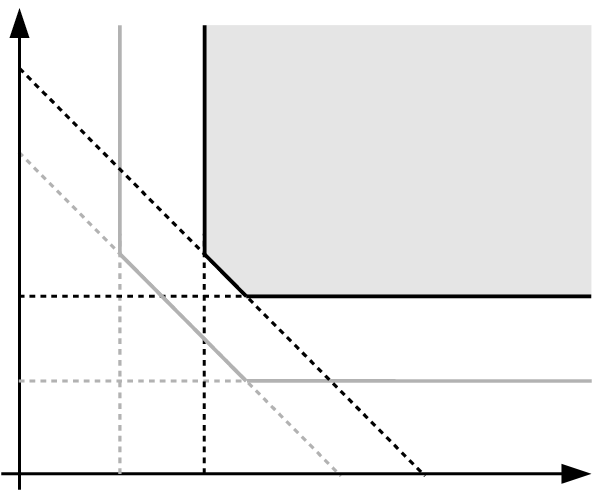}
        \put(0,145){$R_{2}$}
        \put(-10, 120){$R_3^{*}$}
        \put(-35, 95){$H(X, Y)$}
        \put(-10, 55){$R_2^{*}$}
        \put(-35, 30){$H(Y|X)$}
        \put(10, -5){$H(X|Y)$}        
        \put(53, -5){$R_1^{*}$}
        \put(75,-5){$H(X, Y)$}
        \put(120, -5){$R_{3}^{*}$}
        \put(175, 0){$R_{1}$}        

        \put(100, 90){$\cR_{\bfd}(\cS)$}
    \end{overpic}
    \vspace{5mm}
    \caption{An image of $\cR_{\bfd}(\cS)$, where $P_{XY} \in \cS$, $R_1^{*} = \sup_{P_{XY}\in\cS} \left(H(X|Y) + \Delta_{\bfd} I(X;Y)\right)$, $R_2^{*} = \sup_{P_{XY}\in\cS} \left(H(Y|X) + \Delta_{\bfd} I(X;Y)\right)$, and $R_3^{*} = \sup_{P_{XY}\in\cS} \left(H(X, Y) + \Delta_{\bfd} I(X;Y)\right)$.}
    \label{fig:rate_region}
\end{figure*}

The next theorem clarifies the achievable rate region (see also Fig.\ \ref{fig:rate_region}).
\begin{thm}
    \label{thm:acievable rate region for asy SW}
    For a set $\cS$ of PMFs of DMSs and a bound $\bfd$ of delays, we have
    \begin{align*}
        \cR_{\bfd}(\cS) &= \left\{(R_1,R_2): R_i \geq \sup_{P_{XY}\in\cS} \left(H_{i}(X, Y) + \Delta_{\bfd} I(X;Y)\right),\ \forall i \in \{1, 2, 3\} \right\},
    \end{align*}
    where
    \begin{align*}
        \Delta_{\bfd} \teq \limsup_{n\ra\infty} \frac{\max\left\{|\overline d_{n}|, |\underline d_{n}|\right\}}{n}.
    \end{align*}
\end{thm}
\begin{rem}
    By the definition of $\bfd$, it holds that $0\leq \Delta_{\bfd} \leq 1$.
\end{rem}

The proof of this theorem is given later in Sections \ref{sec:converse part} and \ref{sec:direct part}. In this theorem, $\Delta_{\bfd}$ denotes the ratio of the maximum delay to the blocklength. For example, for an arbitrarily fixed constant $\A \in (0, 1]$, suppose that $\underline d_{n}=-\A n$ and $\overline d_{n}=\A n$. Then $\Delta_{\bfd}=\A$, and hence we have for any subset $\cS \subseteq \cP(\cX \times \cY)$,
\begin{align}
    \cR_{\bfd}(\cS) &= \left\{(R_1,R_2): R_i \geq \sup_{P_{XY}\in\cS} \left(H_{i}(X, Y) + \A I(X;Y)\right),\ \forall i \in \{1, 2, 3\} \right\}.
    \label{equ: achievable rate region for alpha}
\end{align}
Since $\alpha > 0$, the rate region does not coincide with $\cR(X, Y)$, i.e., that of the synchronous SW coding system, where we assume that $P_{XY} \in \cS$ and $I(X; Y) > 0$. This means that if delays have a significant influence in terms of the blocklength $n$ like this example, we cannot achieve a pair of rates of the synchronous SW coding system. As shown in a later section, this is because delays affect the \textit{first-order} of rates. Although this fact is very important, it is not clear from previous studies \cite{willems1988totally} and \cite{oki1997coding}.

More generally, we have the following corollaries.
\begin{cor}
    \label{cor:AwithD_not_equal_AwithoutD}
    Let $P_{XY} \in \cS$. Then, $\cR_{\bfd}(\cS) \neq \cR(X, Y)$ if and only if $\sup_{P_{XY}\in\cS} \left(H_{i}(X, Y) + \Delta_{\bfd} I(X;Y)\right) \neq H_{i}(X, Y)$ for some $i \in \{1, 2, 3\}$.
\end{cor}
\begin{IEEEproof}
    If it holds that $\sup_{P_{XY}\in\cS} \left(H_{i}(X, Y) + \Delta_{\bfd} I(X;Y)\right) \neq H_{i}(X, Y)$ for some $i \in \{1, 2, 3\}$, we have
    \begin{align*}
        \sup_{P_{XY}\in\cS} \left(H_{i}(X, Y) + \Delta_{\bfd} I(X;Y)\right) > H_{i}(X, Y).
    \end{align*}
    For such $i \in \{1, 2, 3\}$, we choose a boundary point $(R'_1, R'_2) \in \cR(X, Y)$ such that $R'_i = H_{i}(X, Y)$. Then, according to Theorem \ref{thm:acievable rate region for asy SW}, we have
    \begin{align*}
        \inf\{R_i: (R_1, R_2) \in \cR_{\bfd}(\cS)\} \geq \sup_{P_{XY}\in\cS} \left(H_{i}(X, Y) + \Delta_{\bfd} I(X;Y)\right) > H_{i}(X, Y) = R'_i.
    \end{align*}
    This means that $(R'_1, R'_2) \notin \cR_{\bfd}(\cS)$, and hence $\cR_{\bfd}(\cS) \neq \cR(X, Y)$. This gives the \textit{if} part.

    On the other hand, if it holds that $\sup_{P_{XY}\in\cS} \left(H_{i}(X, Y) + \Delta_{\bfd} I(X;Y)\right) = H_{i}(X, Y)$ for all $i \in \{1, 2, 3\}$, it immediately holds that $\cR_{\bfd}(\cS) = \cR(X, Y)$. This completes the proof.
\end{IEEEproof}
\begin{cor}
    Let $\cS = \{P_{XY}\}$ be a singleton. Then, $\cR_{\bfd}(X, Y) \neq \cR(X, Y)$ if and only if $\Delta_{\bfd} \neq 0$ and $I(X; Y) \neq 0$.
\end{cor}
\begin{IEEEproof}
    According to Corollary \ref{cor:AwithD_not_equal_AwithoutD}, $\cR_{\bfd}(X, Y) \neq \cR(X, Y)$ if and only if
    \begin{align*}
        H_{i}(X, Y) + \Delta_{\bfd} I(X;Y) \neq H_{i}(X, Y),\  \exists i \in \{1, 2, 3\}.
    \end{align*}
    This holds if and only if $\Delta_{\bfd} \neq 0$ and $I(X; Y) \neq 0$.
\end{IEEEproof}

\section{Converse Part}
\label{sec:converse part}
In this section, we give a proof of the converse part of Theorem \ref{thm:acievable rate region for asy SW}.

Before the proof of the converse part, we show the next fundamental lemma.
\begin{lem}
    \label{lem: EntropyWithDelay}
    For any $P_{XY}\in\cS$ and $d\in\cD_n$, we have
    \begin{align}
        H_{i}(X^n, Y^n_{(d)}) &= n H_{i}(X, Y)+|d|I(X;Y),\ \forall i \in \{1, 2, 3\}. \label{eq:CondEntropyX of delay}
    \end{align}
\end{lem}
\begin{IEEEproof}
    According to (\ref{eq:dist. of a source with relative delays}), for any $P_{XY}\in\cS$ and delay $d\in\cD_{n}$, we have
    \begin{align*}
        &H(X^n|Y^n_{(d)})\\
        &= -\sum_{(x^n,y^n)\in\cX^n\times\cY^n}
        P_{X}^{|d|}(x_{\rm I}^{|d|})
        P_{XY}^{n-|d|}(x_{\rm C}^{n-|d|}, y_{\rm C}^{n-|d|})
        P_{Y}^{|d|}(y_{\rm I}^{|d|}) \log
        P_{X}^{|d|}(x_{\rm I}^{|d|})
        P_{X|Y}^{n-|d|}(x_{\rm C}^{n-|d|}|y_{\rm C}^{n-|d|})\\
        &= -\sum_{x^{|d|}\in\cX^{|d|}}
        P_{X}^{|d|}(x^{|d|})
        \log P_{X}^{|d|}(x^{|d|})\\
        &\quad\ -\sum_{(x^{n-|d|}, y^{n-|d|})\in\cX^{n-|d|}\times \cY^{n-|d|}}
        P_{XY}^{n-|d|}(x^{n-|d|}, y^{n-|d|})
        \log P_{X|Y}^{n-|d|}(x^{n-|d|}|y^{n-|d|})\\
        &= |d|H(X) + (n-|d|)H(X|Y)\\
        &= nH(X|Y)+|d|I(X;Y).
    \end{align*}
    Similarly, we have
    \begin{align*}
        H(Y^n_{(d)}|X^n) = nH(Y|X)+|d|I(X;Y),
    \end{align*}
    and
    \begin{flalign*}
        & & H(X^n, Y^n_{(d)})
        & = |d|H(X) + (n-|d|)H(X,Y) + |d|H(Y) \\
        & & & = nH(X, Y)+|d|I(X;Y). & \IEEEQEDhere
    \end{flalign*}
\end{IEEEproof}

In order to emphasize the affect of delays, we give a converse bound for a given finite blocklength in the next theorem.
\begin{thm}
    \label{thm: finite blocklength converse}
    For any $n > 0 $ and any code $(f_n^{(1)}, f_n^{(2)}, \varphi_n)$, we have
    \begin{align*}
        R_n^{(i)} &\geq \sup_{P_{XY}\in\cS} \left(H_{i}(X, Y)+ \Delta_{\bfd}^{(n)} I(X;Y)\right) - \epsilon_{n},\ \forall i \in \{1, 2, 3\},
    \end{align*}
    where $\Delta_{\bfd}^{(n)} = \frac{\max\left\{|\overline d_{n}|, |\underline d_{n}|\right\} }{n}$, $R_n^{(3)} = R_n^{(1)} +  R_n^{(2)}$,
    and
    \begin{align*}
        \epsilon_{n} = \sup_{P_{XY}\in\cS} \max_{d\in\cD_{n}} \vep_{XY_{(d)}}^{(n)} \log |\cX| |\cY| + \frac{1}{n}.
    \end{align*}
\end{thm}
\begin{IEEEproof}
    By using a usual converse technique (see Appendix \ref{sec:converse_bound}), for any $n > 0$, code $(f_n^{(1)}, f_n^{(2)}, \varphi_n)$, delay $d\in\cD_{n}$, and $P_{XY}\in\cS$, we have
    \begin{align}
        nR_n^{(i)} &\geq H_{i}(X^n, Y^n_{(d)}) - n \vep_{XY_{(d)}}^{(n)} \log |\cX| |\cY| - 1,\ \forall i \in \{1, 2, 3\}.
        \label{equ:lower_from_Fano's_ineq}
    \end{align}
    Hence, we have
    \begin{align*}
        R_n^{(i)} &\geq \sup_{P_{XY}\in\cS} \max_{d\in\cD_{n}} \left( \frac{1}{n} H_{i}(X^n, Y^n_{(d)}) - \vep_{XY_{(d)}}^{(n)} \log |\cX| |\cY| - \frac{1}{n} \right)\\
        &\eqo{(a)} \sup_{P_{XY}\in\cS} \max_{d\in\cD_{n}} \left( H_{i}(X, Y) + \frac{|d|}{n} I(X;Y) - \vep_{XY_{(d)}}^{(n)} \log |\cX| |\cY| - \frac{1}{n} \right)\\
        &\geq \sup_{P_{XY}\in\cS} \max_{d\in\cD_{n}} \left( H_{i}(X, Y) + \frac{|d|}{n} I(X;Y) \right) - \left(  \sup_{P_{XY}\in\cS} \max_{d\in\cD_{n}} \vep_{XY_{(d)}}^{(n)} \log |\cX| |\cY| + \frac{1}{n} \right)\\
        &= \sup_{P_{XY}\in\cS} \left( H_{i}(X, Y) + \frac{ \max_{d\in\cD_{n}} |d|}{n} I(X;Y) \right) - \epsilon_{n}\\
        &= \sup_{P_{XY}\in\cS} \left( H_{i}(X, Y) + \Delta_{\bfd}^{(n)} I(X;Y) \right) - \epsilon_{n},\ \forall i \in \{1, 2, 3\},
    \end{align*}
    where (a) comes from Lemma \ref{lem: EntropyWithDelay}.
\end{IEEEproof}
According to this theorem, delay affects the first-order of rates as the term $\Delta_{\bfd}^{(n)} I(X;Y)$. Hence, the larger $\Delta_{\bfd}^{(n)}$ becomes, the greater the difference between rates of synchronous SW coding systems and that of asynchronous SW coding systems. As an instance, we consider the case were delay occurs a ratio of $\alpha > 0$ for a given blocklength $n$, i.e., $\underline d_{n} = - \alpha n$ and $\overline d_{n} = \alpha n$. Since the error probability must be small, this theorem implies that any pair of rates $(R_n^{(1)}, R_n^{(2)})$ must satisfy 
\begin{align*}
    R_n^{(i)} &\gtrsim \sup_{P_{XY}\in\cS} \left(H_{i}(X, Y)+ \alpha I(X;Y)\right),\ \forall i \in \{1, 2, 3\}.
\end{align*}
This also justifies the achievable rate region in Theorem \ref{thm:acievable rate region for asy SW} (cf.\ \eqref{equ: achievable rate region for alpha}).

Now, we prove the converse part.
\begin{IEEEproof}[Proof of the converse part]
    Suppose that $(R_1, R_2) \in \cR_{\bfd}(\cS)$. By definition of the achievability, there exists a sequence of codes $\{(f_n^{(1)}, f_n^{(2)}, \varphi_n)\}$ such that
    \begin{align*}
        \limsup_{n\ra\infty } R_n^{(i)} &\leq R_i,\ \forall i \in \{1, 2\},\\
        \lim_{n\ra\infty} \sup_{P_{XY} \in\cS} \max_{d \in\cD_{n}}
        \vep_{XY_{(d)}}^{(n)}(f_n^{(1)}, f_n^{(2)}, \varphi_n) &= 0.
    \end{align*}
    According to Theorem \ref{thm: finite blocklength converse}, for this sequence of codes and any $P_{XY}\in\cS$, we have
\begin{align*}
    R_{i} &\geq \limsup_{n\ra\infty}R_n^{(i)}\\
    & \geq H_{i}(X, Y)+\Delta_{\bfd} I(X;Y),\ \forall i \in \{1, 2, 3\},
\end{align*}
where we use the fact that
\begin{align*}
    \limsup_{n\ra\infty} \Delta_{\bfd}^{(n)} = \Delta_{\bfd}.
\end{align*}
Since this inequality holds for any $P_{XY}\in\cS$, we have
\begin{align*}
    R_{i} & \geq \sup_{P_{XY}\in\cS} \left(H_{i}(X, Y)+\Delta_{\bfd} I(X;Y)\right),\ \forall i \in \{1, 2, 3\}.
\end{align*}

Since this holds for any $(R_1, R_2) \in \cR_{\bfd}(\cS)$, we have
\begin{align*}
    \cR_{\bfd}(\cS) &\subseteq \left\{(R_1,R_2): R_i \geq \sup_{P_{XY}\in\cS} \left(H_{i}(X, Y) + \Delta_{\bfd} I(X;Y)\right),\ \forall i \in \{1, 2, 3\} \right\}.
\end{align*}
This completes the proof of the converse part of Theorem \ref{thm:acievable rate region for asy SW}.
\end{IEEEproof}

\section{Direct Part and a Universal Coding Scheme}
\label{sec:direct part}
In this section, we give a proof of the direct part of Theorem \ref{thm:acievable rate region for asy SW}. To this end, we will show a universal coding scheme for asynchronous SW coding systems.

The universal coding scheme using the minimum entropy decoder \cite{csiszar1982lcs} is well known. However, since it is too much specialized to DMSs without delay, we cannot use it to our coding system straightforwardly. A natural extension of the minimum entropy decoder may be as follows:
\begin{align*}
    \varphi_{n}(m_1,m_2)=
    \begin{cases}
        (x^n,y^n) &\mbox{ if } (f_{n}^{(1)}(x^n),f_{n}^{(2)}(y^n)) = (m_1,m_2),\\
        & \exists d\in\cD_{n},\ \forall \hat d\in\cD_{n},\ \forall (\tilde x^n,\tilde y^n) \in \cX^n \times \cY^n, \mbox{ s.t. }\\
        & (f_{n}^{(1)}(\tilde x^n),f_{n}^{(2)}(\tilde y^n)) = (m_1,m_2),\\
        & |d|H(P_{x_{\rm I}^{|d|}}) + (n-|d|)H(P_{(x_{\rm C}^{n-|d|}, y_{\rm C}^{n-|d|})}) + |d|H(P_{y_{\rm I}^{|d|}})\\
        & < |\hat d|H(P_{\tilde x_{\rm I}^{|\hat d|}}) + (n-|\hat d|)H(P_{(\tilde x_{\rm C}^{n-|\hat d|}, \tilde y_{\rm C}^{n-|\hat d|})}) + |\hat d|H(P_{\tilde y_{\rm I}^{|\hat d|}}),\\
        $\mbox {declare the error}$ &\mbox{ otherwise},
    \end{cases}
\end{align*}
where $P_{a^k}$ is the \textit{type} (cf.\ \cite{csiszar2011itc}) of a sequence $a^k$ defined by $P_{a^k}(a) \teq |\{i \in \{1, \cdots, k\}: a_{i} = a\}|/k$. However, it may not be able to evaluate well the error probability of this decoder by using a usual method of the types. This is because, for sequences $y^n$ and $\tilde y^n$ with the same type (or even for $y^n = \tilde y^n$), the type of $y^{n-d}$ does not always coincide with the type of $\tilde y^{n-\hat d}$ when $d\neq \hat d$. Thus, when we use the minimum entropy decoder, we will probably need a more sophisticated and somewhat complicated technique.

Instead of using the minimum entropy decoder, we employ a decoder for a mixed source and analyze the error probability by using Gallager's random coding techniques \cite{gallager1968information} and \cite{gallager1976}. To this end, we introduce $\cP_n(\cX)$ that is the set of PMFs $P_{\bar{X}}\in\cP(\cX)$ taking rational numbers such that for each $x \in \cX$,
\begin{align*}
    P_{\bar{X}}(x)=\frac{m}{n},
\end{align*}
where $m \in \{0, 1, \cdots, n\}$. This type of PMF is referred to as \textit{$n$-type} in this paper. The next bound is well known (see e.g., \cite[Lemma 2.2]{csiszar2011itc}).
\begin{align}
    |\cP_n(\cX)| \leq (n + 1)^{|\cX|}. \label{equ:bound of type}
\end{align}

Let $d_{\rm v}$ be the variational distance between two PMFs $P, Q \in \cP(\cX)$ defined as
\begin{align*}
    d_{\rm v}(P, Q) \teq \sum_{x\in\cX} |P(x) - Q(x)|.
\end{align*}
Then, we have the next lemma which gives an approximation of a given distribution by an $n$-type. The proof is given in Appendix \ref{sec:proof of lem: exists of type for delay}.
\begin{lem}
    \label{lem: exists of type for delay}
    Let $\alpha = |\cX| |\cY|(|\cX| |\cY|-1)$ and $\vep_n = \A \ln \big(1 + \frac{\A}{n - \A}\big)$. Then, for any $n\geq \alpha$ and $P_{XY} \in \cP(\cX\times\cY)$,  there exists an $n$-type $P_{\bar{X}\bar{Y}}\in \cP_{n}(\cX\times\cY)$ such that
    \begin{align}
        d_{\rm v}(P_{XY}, P_{\bar{X}\bar{Y}}) \leq \frac{2(|\cX| |\cY|-1)}{n},
        \label{lem: equ: variational distance of prob and type}
    \end{align}
    and
    \begin{align}
        P_{X^{n}Y_{(d)}^{n}}(x^n,y^n) & \leq  P_{\bar X^{n}\bar Y_{(d)}^{n}}(x^n,y^n)e^{3(\A+\vep_n)},\notag\\
        &\quad \forall ({x^n},{y^n})\in \cX^n \times \cY^n,\ \forall d \in \{-n, \cdots, n\},
        \label{lem: eqn: n-fold bound by the type with delay}
    \end{align}
    where $e$ is the base of the natural logarithm.
\end{lem}

For each $P_{XY} \in \cS$, we consider the corresponding $n$-type $P_{\bar X \bar Y}$ in Lemma \ref{lem: exists of type for delay}. We denote the set of those $n$-types for elements of $\cS$ by $\bar{\cS}_{n} (\subseteq \cP_{n}(\cX \times \cY))$, where $\bar{\cS}_{n}$ is arbitrary for $n < \alpha$. Hence by definition, for any $P_{XY} \in \cS$ (and $n \geq \alpha$), there exists an $n$-type $P_{\bar X \bar Y} \in \bar{\cS}_{n}$ satisfying \eqref{lem: equ: variational distance of prob and type} and \eqref{lem: eqn: n-fold bound by the type with delay}. Roughly speaking, $\bar{\cS}_{n}$ is an approximation of $\cS$ by $n$-types. By using this set $\bar{\cS}_{n}$, we  define the \textit{mixed source} $(\widetilde{\bfX},\widetilde{\bfY}) =\{(\widetilde{X}^n,\widetilde{Y}^n)\}_{n=1}^{\infty}$,
where the pair of RVs $(\widetilde{X}^n,\widetilde{Y}^n)$ is drawn according to the joint PMF $P_{\widetilde{X}^n\widetilde{Y}^n}$ defined as
\begin{align}
    P_{\widetilde{X}^n\widetilde{Y}^n}(x^n, y^n)
    \teq& \sum_{P_{\bar{X}\bar{Y}}\in \bar{\cS}_{n}} \sum_{d\in \cD_n} \frac{1}{|\bar{\cS}_{n}||\cD_n|} P_{\barX^{n}\barY_{(d)}^{n}}(x^n,y^n). \label{equ:P_of_mixed_source}
\end{align}

This mixed source gives a bound on the maximum error probability for DMSs with delay as shown in the next theorem.
\begin{thm}
    \label{thm: cond of universal code}
    Let $\alpha = |\cX| |\cY|(|\cX| |\cY|-1)$ and $\vep_n = \A \ln \big(1 + \frac{\A}{n - \A}\big)$. Then, for any $n \geq \alpha$ and any code $(f_n^{(1)}, f_n^{(2)}, \varphi_n)$, we have
    \begin{align}
        &\sup_{P_{X Y} \in \cS} \max_{d \in \cD_{n}} \vep_{XY_{(d)}}^{(n)}(f_n^{(1)}, f_n^{(2)}, \varphi_n) \leq e^{3(\A+\vep_n)} (2 n + 1) (n + 1)^{|\cX| |\cY|} \vep_{\widetilde{\bf X}\widetilde{\bf Y}}^{(n)}(f_n^{(1)}, f_n^{(2)}, \varphi_n).
        \label{equ: thm: cond of universal code}
    \end{align}
\end{thm}

\begin{IEEEproof}
    We prove this theorem in two parts. In the first part, we show that for any $n > 0$ and any code $(f_n^{(1)}, f_n^{(2)}, \varphi_n)$,
    \begin{align}
        &\max_{P_{\bar{X}\bar{Y}}\in \bar{\cS}_{n}} \max_{d\in\cD_{n}} \vep_{\bar{X}\bar {Y}_{(d)}}^{(n)}(f_n^{(1)}, f_n^{(2)}, \varphi_n) \leq (2 n + 1) (n + 1)^{|\cX| |\cY|} \vep_{\widetilde{\bf X}\widetilde{\bf Y}}^{(n)}(f_n^{(1)}, f_n^{(2)}, \varphi_n),
        \label{equ: first part universal codeing}
    \end{align}
    where $(\bar X,\bar Y_{(d)})$ denotes DMSs $(\bar X, \bar Y)$ induced by the $n$-type $P_{\bar X \bar Y}$ with the delay $d$. In the second part, we show that for any $n \geq \alpha$ and any code $(f_n^{(1)}, f_n^{(2)}, \varphi_n)$,
    \begin{align}
        & \sup_{P_{XY} \in \cS} \max_{d \in \cD_{n}} \vep_{XY_{(d)}}^{(n)}(f_n^{(1)}, f_n^{(2)}, \varphi_n) \leq e^{3 (\A + \vep_n)} \max_{P_{\bar{X} \bar{Y}} \in \bar{\cS}_{n}} \max_{d \in \cD_{n}} \vep_{\bar{X} \bar{Y}_{(d)}}^{(n)}(f_n^{(1)}, f_n^{(2)}, \varphi_n).
        \label{equ: second part universal codeing}
    \end{align}
    The theorem follows immediately from \eqref{equ: first part universal codeing} and \eqref{equ: second part universal codeing}.

    \textbf{First part:}
    Since the decoder $\varphi_n$ is a deterministic function, for any source $(\bfX,\bfY)$,
    the error probability $\vep_{\bfX\bfY}^{(n)}(f_n^{(1)}, f_n^{(2)}, \varphi_n)$ can be written as
    \begin{align*}
        \vep_{\bfX\bfY}^{(n)}(f_n^{(1)}, f_n^{(2)}, \varphi_n)
        = \sum_{(x^n, y^n)\in \cE_n(f_n^{(1)}, f_n^{(2)}, \varphi_n)} P_{X^nY^{n}}(x^n,y^n),
    \end{align*} 
    where $\cE_n(f_n^{(1)}, f_n^{(2)}, \varphi_n)$ is the set of pairs of sequences which cannot be decoded correctly, i.e.,
    \begin{align*}
        \cE_n(f_n^{(1)}, f_n^{(2)}, \varphi_n)
        \teq \{(x^n, y^n)\in\cX^n\times\cY^n: \varphi_n(f_n^{(1)}(x^n),f_n^{(2)}(y^n)) \neq (x^n, y^n)\}.
    \end{align*}
    Thus, we have
    \begin{align*}
        \max_{P_{\bar{X}\bar{Y}}\in \bar{\cS}_{n}}\max_{d\in\cD_{n}} \vep_{\bar X\bar Y_{(d)}}^{(n)}(f_n^{(1)}, f_n^{(2)}, \varphi_n)
        &\leq \sum_{P_{\bar{X}\bar{Y}}\in \bar{\cS}_{n}} \sum_{d\in\cD_n}
        \vep_{\bar X\bar Y_{(d)}}^{(n)}(f_n^{(1)}, f_n^{(2)}, \varphi_n)\\
        &=\sum_{P_{\bar{X}\bar{Y}}\in \bar{\cS}_{n}}
        \sum_{d\in\cD_n} \sum_{(x^n, y^n)\in \cE_n(f_n^{(1)}, f_n^{(2)}, \varphi_n)} P_{\barX^{n}\barY_{(d)}^{n}}(x^n,y^n)\\
        &= |\cD_n| |\bar{\cS}_{n}| \sum_{(x^n, y^n)\in \cE_n(f_n^{(1)}, f_n^{(2)}, \varphi_n)}
        P_{\widetilde{X}^n\widetilde{Y}^n}(x^n, y^n)\\
        &= |\cD_n| |\bar{\cS}_{n}|\vep_{\widetilde{\bf X}\widetilde{\bf Y}}^{(n)}(f_n^{(1)}, f_n^{(2)}, \varphi_n)\\
        &\leq (2 n + 1) (n + 1)^{|\cX| |\cY|}
        \vep_{\widetilde{\bf X}\widetilde{\bf Y}}^{(n)}(f_n^{(1)}, f_n^{(2)}, \varphi_n),
    \end{align*}
    where the last inequality comes from the fact that  $-n\leq \underline{d}_{n} \leq \overline{d}_{n}\leq n$ and (\ref{equ:bound of type}). Hence, we have \eqref{equ: first part universal codeing}.

    {\bf Second part:} For an arbitrarily fixed $P_{XY} \in \cS$, we consider a corresponding $n$-type $P_{\bar X \bar Y} \in \bar{\cS}_{n}$ satisfying \eqref{lem: eqn: n-fold bound by the type with delay} in Lemma \ref{lem: exists of type for delay}. Then, for any $d \in \cD_{n}$, $\vep_{XY_{(d)}}^{(n)}(f_n^{(1)}, f_n^{(2)}, \varphi_n)$ can be bounded by
    \begin{align*}
        \vep_{XY_{(d)}}^{(n)}(f_n^{(1)}, f_n^{(2)}, \varphi_n)
        & = \sum_{(x^n, y^n)\in \cE_n(f_n^{(1)}, f_n^{(2)}, \varphi_n)} P_{X^n Y_{(d)}^n}(x^n, y^n)\nonumber \\
        & \leq  e^{3(\A+\vep_n)} \sum_{(x^n, y^n)\in \cE_n(f_n^{(1)}, f_n^{(2)}, \varphi_n)} P_{\barX^n\barY_{(d)}^n}(x^n, y^n) \nonumber \\
        & =  e^{3(\A+\vep_n)} \vep_{\bar X \bar Y_{(d)}}^{(n)}(f_n^{(1)}, f_n^{(2)}, \varphi_n) \\
        & \leq e^{3(\A+\vep_n)} \max_{P_{\bar{X} \bar{Y}} \in \bar{\cS}_{n}} \max_{d \in \cD_{n}} \vep_{\bar X \bar Y_{(d)}}^{(n)}(f_n^{(1)}, f_n^{(2)}, \varphi_n),
    \end{align*}
    where the first inequality follows since $P_{\bar X \bar Y}$ satisfies \eqref{lem: eqn: n-fold bound by the type with delay}. This inequality holds for any $P_{XY} \in \cS$ and $d \in \cD_{n}$, we have \eqref{equ: second part universal codeing}.
\end{IEEEproof}

This theorem implies that if the error probability for the mixed source vanishes at the order $o(n^{-|\cX| |\cY|-1})$, the maximum error probability for DMSs with delay also vanishes. In the next theorem, we show the existence of a code of which the error probability vanishes exponentially rather than $o(n^{-|\cX| |\cY|-1})$.

To this end, we introduce some notations: For $P_{XY}\in\cP(\cX\times\cY)$ and a real number $\rho\geq 0$, let $P_{X}^{(\rho)}\in\cP(\cX)$ be a PMF given by
\begin{align}
    P_{X}^{(\rho)}(x) &= \frac{P_{X}(x)^{\frac{1}{1 + \rho}}}{\sum_{x\in\cX}P_{X}(x)^{\frac{1}{1 + \rho}}}, \label{equ: PXrho}
\end{align}
$P_{X|Y}^{(\rho)}\in\cP(\cX|\cY)$ be a conditional PMF given by
\begin{align}
    P_{X|Y}^{(\rho)}(x|y) &= \frac{P_{XY}(x,y)^{\frac{1}{1 + \rho}}} {\sum_{x\in\cX} P_{XY}(x,y)^{\frac{1}{1 + \rho}}}, \label{equ: barPX|Yrho}
\end{align}
and $\bar P_{X}^{(\rho)}\in\cP(\cX)$ be a PMF given by
\begin{align}
    \bar P_{X}^{(\rho)}(x) &= \frac{\left(\sum_{y\in\cY} P_{XY}(x,y)^{\frac{1}{1 + \rho}}\right)^{1 + \rho}}{\sum_{x\in\cX}\left(\sum_{y\in\cY} P_{XY}(x,y)^{\frac{1}{1 + \rho}}\right)^{1 + \rho}}. \label{equ: barPXrho}
\end{align}
We note that these PMFs are continuous with respect to $\rho \in [0, 1]$. We also note that when $\rho = 0$, it holds that $P_{X}^{(\rho)} = \bar P_{X}^{(\rho)} =P_{X}$ and $P_{X|Y}^{(\rho)} = P_{X|Y}$.

Now, we show an exponential bound on the error probability.
\begin{thm}
    \rm\label{thm:existence of mixed code}
    For any $R_1, R_2 > 0$ and $n \geq \alpha$, there exists a code $(f_n^{(1)}, f_n^{(2)}, \varphi_n)$ such that
    \begin{align*}
        M_n^{(1)} = \left\lceil 2^{n R_1}\right\rceil,\
        M_n^{(2)} = \left\lceil 2^{n R_2}\right\rceil,
    \end{align*}
    and
    \begin{align*}
        -\frac{1}{n}\log \vep_{\widetilde{\bf X}\widetilde{\bf Y}}^{(n)}
        \geq \min_{i\in\{1,2,3\}} \sup_{0\leq \rho \leq 1}\min_{P_{XY}\in\bar{\cS}_{n}} F_{i}(\rho, R_i, P_{XY},\Delta_{\bfd}^{(n)}) - \epsilon_{n},
    \end{align*}
    where $\epsilon_{n} = \frac{\log3(2 n + 1) (n + 1)^{|\cX| |\cY|}}{n}$,
    \begin{align}
        F_{1}(\rho, R_1, P_{XY},\Delta) &= \rho \left(R_1 - H(P_{X|Y}^{(\rho)}|\bar P_{Y}^{(\rho)}) - \Delta\left| H(P_{X}^{(\rho)}) - H(P_{X|Y}^{(\rho)}|\bar P_{Y}^{(\rho)}) \right|_{+} \right)\notag\\
        &\quad + D(\bar P_{Y}^{(\rho)}\times P_{X|Y}^{(\rho)}\|P_{XY}) - \Delta \left|D(\bar P_{Y}^{(\rho)}\times P_{X|Y}^{(\rho)}\|P_{XY}) - D(P_{X}^{(\rho)}\|P_{X}) \right|_{+},
        \label{eq: ExFunction1}\\
        F_{2}(\rho, R_2, P_{XY},\Delta) &= \rho \left(R_2 - H(P_{Y|X}^{(\rho)}|\bar P_{X}^{(\rho)}) - \Delta\left| H(P_{Y}^{(\rho)}) - H(P_{Y|X}^{(\rho)}|\bar P_{X}^{(\rho)}) \right|_{+} \right)\notag\\
        &\quad + D(\bar P_{X}^{(\rho)}\times P_{Y|X}^{(\rho)}\|P_{XY}) - \Delta\left| D(\bar P_{X}^{(\rho)}\times P_{Y|X}^{(\rho)}\|P_{XY}) - D(P_{Y}^{(\rho)}\|P_{Y})\right|_{+},
        \label{eq: ExFunction2}\\
        F_{3}(\rho,R_3,P_{XY},\Delta) & = \rho \left( R_3 - H(P_{XY}^{(\rho)}) - \Delta \left|H(P_{X}^{(\rho)}) + H(P_{Y}^{(\rho)}) - H(P_{XY}^{(\rho)})\right|_{+} \right)\notag\\
        &\quad + D(P_{XY}^{(\rho)}\|P_{XY}) - \Delta\left| D(P_{XY}^{(\rho)}\|P_{XY}) - D(P_{X}^{(\rho)}\|P_{X}) - D(P_{Y}^{(\rho)}\|P_{Y}) \right|_{+}, \label{eq: ExFunction3}
    \end{align}
    and $|x|_{+}\teq\max\{0,x\}$.
\end{thm}

\begin{IEEEproof}
    By employing Gallager's random coding techniques \cite{gallager1968information} and \cite{gallager1976} to the mixed source $(\widetilde{\bf X}, \widetilde{\bf Y})$, we can show the existence of a code $(f_n^{(1)}, f_n^{(2)}, \varphi_n)$ such that $M_n^{(1)} = \left\lceil 2^{n R_1}\right\rceil$, $M_n^{(2)} = \left\lceil 2^{n R_2}\right\rceil$, and
    \begin{align}
        \vep_{\widetilde{\bf X}\widetilde{\bf Y}}^{(n)}(f_n^{(1)}, f_n^{(2)}, \varphi_n) &\leq  3 \exp\left( -n \min_{i \in \{1, 2, 3\}} \max_{0\leq\rho\leq 1} \left(\rho R_i-E_n^{(i)}(\rho,P_{\widetilde{X}^n\widetilde{Y}^n})\right) \right),
        \label{inq:proof:GallagerBound}
    \end{align}
    where
    \begin{align}
        E_n^{(1)}(\rho,P_{X^nY^n}) 
        &= \frac{1}{n}\log \sum_{y^n\in\cY^n}
        \left(\sum_{x^n\in\cX^n}P_{X^nY^n}(x^n,y^n)^{\frac{1}{1 + \rho}}\right)^{1 + \rho},
        \label{equ: En1}\\ 
        E_n^{(2)}(\rho,P_{X^nY^n}) 
        &= \frac{1}{n}\log\sum_{x^n\in\cX^n}
        \left(\sum_{y^n\in\cY^n}P_{X^nY^n}(x^n,y^n)^{\frac{1}{1 + \rho}}\right)^{1 + \rho},
        \label{equ: En2}\\ 
        E_n^{(3)}(\rho,P_{X^nY^n})
        &= \frac{1}{n}\log
        \left(\sum_{(x^n,y^n)\in\cX^n\times\cY^n}
        P_{X^nY^n}(x^n,y^n)^{\frac{1}{1 + \rho}}\right)^{1 + \rho}.
        \label{equ: En3}
    \end{align}
    This bound is also given by Nomura and Han \cite[Appendix C]{nomura2014second} more explicitly.
    
    On the other hand, according to Appendix \ref{appendix: GTBMS}, we have
    \begin{align}
        \max_{0\leq\rho\leq 1} \left(\rho R_i - E_n^{(i)}(\rho, P_{\widetilde{X}^n\widetilde{Y}^n})\right) & \geq \sup_{0\leq \rho \leq 1}\min_{P_{XY}\in\bar{\cS}_{n}} F_{i}(\rho, R_i, P_{XY}, \Delta_{\bfd}^{(n)})  - \epsilon'_{n},\ \forall i \in \{1, 2, 3\},
        \label{eq:ExBound1}
    \end{align}
    where
    \begin{align*}
        \epsilon'_{n} = \frac{\log(2 n + 1) (n + 1)^{|\cX| |\cY|}}{n}.
    \end{align*}
    By combining \eqref{inq:proof:GallagerBound} and \eqref{eq:ExBound1}, we have the theorem.
\end{IEEEproof}

By using Theorems \ref{thm: cond of universal code} and \ref{thm:existence of mixed code}, we obtain the next theorem which gives a bound on the error exponent for the asynchronous SW coding system.
\begin{thm}
    \rm\label{thm:existence of seq of mixed codes}
    For any $R_1, R_2>0$, there exists a sequence of codes $\{(f_n^{(1)}, f_n^{(2)}, \varphi_n)\}$ such that
    \begin{align*}
        M_n^{(1)} = \left\lceil 2^{n R_1}\right\rceil,\
        M_n^{(2)} = \left\lceil 2^{n R_2}\right\rceil,\
        \forall n>0,
    \end{align*}
    and
    \begin{align*}
        \liminf_{n\ra\infty}-\frac{1}{n} \log \sup_{P_{X Y} \in \cS} \max_{d \in \cD_{n}} \vep_{XY_{(d)}}^{(n)}
        \geq \sup_{\delta > 0} \min_{i\in\{1,2,3\}} \sup_{0\leq \rho \leq 1}\inf_{P_{XY}\in\cS_{\delta}} F_{i}(\rho, R_i, P_{XY}, \Delta_{\bfd}),
    \end{align*}
    where 
\begin{align*}
    \cS_{\delta} = \bigcup_{P_{XY}\in\cS} \left\{ P_{X'Y'} \in \cP(\cX \times \cY): d_{\rm v}(P_{XY},P_{X'Y'}) \leq \delta \right\}.
\end{align*}
\end{thm}
\begin{IEEEproof}
    For any $n$-type $P_{\bar X \bar Y} \in \bar{\cS}_{n}$, there exists $P_{XY} \in \cS$ satisfying \eqref{lem: equ: variational distance of prob and type} by definition. Thus, for any $\delta > 0$ and $n \geq \frac{2(|\cX| |\cY|-1)}{\delta}$, we have
    \begin{align*}
        d_{\rm v}(P_{XY}, P_{\bar{X}\bar{Y}}) \leq \frac{2(|\cX| |\cY|-1)}{n} \leq \delta.
    \end{align*}
    This implies that
    \begin{align*}
        P_{\bar X \bar Y} \in \left\{ P_{X'Y'} \in \cP(\cX \times \cY): d_{\rm v}(P_{XY},P_{X'Y'}) \leq \delta \right\}
        \subseteq \cS_{\delta}.
    \end{align*}
    Since this holds for any $n$-type $P_{\bar X \bar Y} \in \bar{\cS}_{n}$, we have $\bar{\cS}_{n} \subseteq \cS_{\delta}$ for any $\delta > 0$ and $n > \frac{2(|\cX| |\cY|-1)}{\delta}$.

    According to this property and Theorem \ref{thm:existence of mixed code}, for any $R_1, R_2>0$, there exists a sequence of codes $\{(f_n^{(1)}, f_n^{(2)}, \varphi_n)\}$ such that
    \begin{align*}
        M_n^{(1)} = \left\lceil 2^{n R_1}\right\rceil,\
        M_n^{(2)} = \left\lceil 2^{n R_2}\right\rceil,\
        \forall n>0,
    \end{align*}
    and
    \begin{align}
        &\liminf_{n\ra\infty}-\frac{1}{n} \log \vep_{\widetilde{\bf X}\widetilde{\bf Y}}^{(n)}(f_n^{(1)}, f_n^{(2)}, \varphi_n) \notag\\
        &\geq \liminf_{n\ra\infty} \min_{i\in\{1,2,3\}} \sup_{0\leq \rho \leq 1}\min_{P_{XY}\in\bar{\cS}_{n}} F_{i}(\rho, R_i, P_{XY},\Delta_{\bfd}^{(n)})\notag\\
        &\geq \liminf_{n\ra\infty} \min_{i\in\{1,2,3\}} \sup_{0\leq \rho \leq 1} \inf_{P_{XY}\in\cS_{\delta}} F_{i}(\rho, R_i, P_{XY},\Delta_{\bfd}^{(n)}), \ \forall \delta > 0,
        \label{equ:ex_bound}
    \end{align}
    where the last inequality comes from the fact that $\bar{\cS}_{n} \subseteq \cS_{\delta}$ for any $\delta > 0$ and sufficiently large $n > 0$.

    Furthermore, for this sequence of codes, we have
        \begin{align}
        &\liminf_{n\ra\infty}-\frac{1}{n} \log \sup_{P_{X Y} \in \cS} \max_{d \in \cD_{n}} \vep_{XY_{(d)}}^{(n)}(f_n^{(1)}, f_n^{(2)}, \varphi_n)\notag\\
        &\geqo{(a)} \liminf_{n\ra\infty}-\frac{1}{n} \log e^{3(\A+\vep_n)} (2 n + 1) (n + 1)^{|\cX| |\cY|} \vep_{\widetilde{\bf X}\widetilde{\bf Y}}^{(n)}(f_n^{(1)}, f_n^{(2)}, \varphi_n)\notag\\
        &= \liminf_{n\ra\infty}-\frac{1}{n} \log \vep_{\widetilde{\bf X}\widetilde{\bf Y}}^{(n)}(f_n^{(1)}, f_n^{(2)}, \varphi_n)\notag\\        
        &\geqo{(b)} \sup_{\delta > 0}\liminf_{n\ra\infty} \min_{i\in\{1,2,3\}} \sup_{0\leq \rho \leq 1}\inf_{P_{XY}\in\cS_{\delta}} F_{i}(\rho, R_i, P_{XY},\Delta_{\bfd}^{(n)}),
        \label{equ:ex_bound2}
    \end{align}
    where (a) comes from Theorem \ref{thm: cond of universal code} and (b) comes from \eqref{equ:ex_bound}.
    
    On the other hand, we have
    \begin{align*}
        \left|H(P_{X}^{(\rho)}) - H(P_{X|Y}^{(\rho)}|\bar P_{Y}^{(\rho)}) \right|_{+} 
        &\leqo{\mathrm{(a)}} \left|H(P_{X}^{(\rho)}) \right|_{+}\\ 
        &\leq \log |\cX|,
    \end{align*}
    and
    \begin{align*}
        \left|D(\bar P_{Y}^{(\rho)}\times P_{X|Y}^{(\rho)}\|P_{XY}) - D(P_{X}^{(\rho)}\|P_{X}) \right|_{+} 
        & \leq \left|D(\bar P_{Y}^{(\rho)}\times P_{X|Y}^{(\rho)}\|P_{XY}) \right|_{+} \\
        & \leqo{(b)} \log |\cX|,
    \end{align*}
    where (a) follows since $|\cdot |_{+}$ is a monotonically increasing function, and (b) comes from Corollary \ref{cor: upper bound of D} in Appendix \ref{appendix: GTBMS}. 
    Similarly, we have
    \begin{align*}
        \left| H(P_{Y}^{(\rho)}) - H(P_{Y|X}^{(\rho)}|\bar P_{X}^{(\rho)}) \right|_{+} &\leq \log |\cY|,\\
        \left| D(\bar P_{X}^{(\rho)}\times P_{Y|X}^{(\rho)}\|P_{XY}) - D(P_{Y}^{(\rho)}\|P_{Y})\right|_{+} &\leq \log|\cY|,\\
        \left|H(P_{X}^{(\rho)}) + H(P_{Y}^{(\rho)}) - H(P_{XY}^{(\rho)})\right|_{+} &\leq \log|\cX| |\cY|,\\
        \left| D(P_{XY}^{(\rho)}\|P_{XY}) - D(P_{X}^{(\rho)}\|P_{X}) - D(P_{Y}^{(\rho)}\|P_{Y}) \right|_{+} &\leq \log |\cX| |\cY|.
    \end{align*}
    Since there exists a sequence $\{\epsilon_{n}\}$ such that $\epsilon_{n} > 0$, $\lim_{n\ra\infty}\epsilon_{n} = 0$, and 
    \begin{align*}
        \Delta_{\bfd}^{(n)} \leq \Delta_{\bfd} + \epsilon_{n},\ \forall n > 0,
    \end{align*}
    we have
    \begin{align}
        &F_{1}(\rho, R_1, P_{XY},\Delta_{\bfd}^{(n)})\notag\\
        &= \rho \left(R_1 - H(P_{X|Y}^{(\rho)}|\bar P_{Y}^{(\rho)}) - \Delta_{\bfd}^{(n)}\left| H(P_{X}^{(\rho)}) - H(P_{X|Y}^{(\rho)}|\bar P_{Y}^{(\rho)}) \right|_{+} \right)\notag\notag\\
        &\quad + D(\bar P_{Y}^{(\rho)}\times P_{X|Y}^{(\rho)}\|P_{XY}) - \Delta_{\bfd}^{(n)} \left|D(\bar P_{Y}^{(\rho)}\times P_{X|Y}^{(\rho)}\|P_{XY}) - D(P_{X}^{(\rho)}\|P_{X}) \right|_{+}\notag\\
        &\geq \rho \left(R_1 - H(P_{X|Y}^{(\rho)}|\bar P_{Y}^{(\rho)}) - \Delta_{\bfd} \left| H(P_{X}^{(\rho)}) - H(P_{X|Y}^{(\rho)}|\bar P_{Y}^{(\rho)}) \right|_{+} \right) - \epsilon_{n} \log |\cX| \notag \notag\\
        &\quad + D(\bar P_{Y}^{(\rho)}\times P_{X|Y}^{(\rho)}\|P_{XY}) - \Delta_{\bfd} \left|D(\bar P_{Y}^{(\rho)}\times P_{X|Y}^{(\rho)}\|P_{XY}) - D(P_{X}^{(\rho)}\|P_{X}) \right|_{+}  - \epsilon_{n} \log |\cX| \notag\\
        &= F_{1}(\rho, R_1, P_{XY},\Delta_{\bfd}) - 2\epsilon_{n} \log |\cX|.
        \label{equ:cad_bound_for_ex_bound1}
    \end{align}
    Similarly, we have
    \begin{align}
        F_{2}(\rho, R_2, P_{XY},\Delta_{\bfd}^{(n)}) &\geq  F_{2}(\rho, R_2, P_{XY},\Delta_{\bfd}) - 2 \epsilon_{n} \log |\cY|, \label{equ:cad_bound_for_ex_bound2}\\
        F_{3}(\rho,R_3,P_{XY},\Delta_{\bfd}^{(n)}) &\geq  F_{3}(\rho, R_3, P_{XY},\Delta_{\bfd}) - 2 \epsilon_{n} \log |\cX| |\cY|. \label{equ:cad_bound_for_ex_bound3}
    \end{align} 
    
    Substituting \eqref{equ:cad_bound_for_ex_bound1}--\eqref{equ:cad_bound_for_ex_bound3} into \eqref{equ:ex_bound2}, we have the desired bound.
\end{IEEEproof}

\begin{rem}
    In the proof of Theorem \ref{thm:existence of mixed code}, the existence of a code is shown by using random binning and the maximum likelihood (ML) decoder (see, e.g., \cite[Appendix C]{nomura2014second}). Thus, obtained encoders of Theorem \ref{thm:existence of seq of mixed codes} do not use a set $\cS$ of PMFs and a bound $\bfd$ of delays. On the other hand, the ML decoder must use $\cS$ and $\bfd$ (or the PMF of the mixed source) to calculate probabilities of source sequences. This claim is the same as \cite{oki1997coding}. However, it is not guaranteed that the error exponent for these encoders is bounded by the same exponent in Theorem \ref{thm:existence of seq of mixed codes} for other parameters of $\cS$ and $\bfd$. Thus, in this theorem, we do not state that the encoders are independent of $\cS$ and $\bfd$. In Section \ref{sec:elimination_knowledge}, we give a sequence of codes that are independent of $\bfd$ and whose error exponent is bounded by the same exponent in Theorem \ref{thm:existence of seq of mixed codes} if delays are bounded.
\end{rem}

We are interested in the condition that the error exponent of Theorem \ref{thm:existence of seq of mixed codes} is positive. The next theorem shows that it is positive whenever $(R_1,R_2)$ is in the achievable rate region. The proof of the theorem is given in Appendix \ref{sec:proof of error exponent}.
\begin{thm}
    \label{thm: positivity of E}
    For any $\cS \subseteq \cP(\cX \times \cY)$, $R_1, R_2 > 0$, and $\Delta \in [0, 1]$ such that
    \begin{align*}
        R_i > \sup_{P_{XY}\in\cS} \left(H_{i}(X, Y)+\Delta I(X;Y)\right),\ \forall i \in \{1, 2, 3\},
    \end{align*}
    we have
    \begin{align}
        \sup_{\delta > 0} \min_{i\in\{1,2,3\}} \sup_{0\leq \rho \leq 1} \inf_{P_{XY}\in\cS_{\delta}} F_{i}(\rho, R_i, P_{XY}, \Delta) > 0.
        \label{thm: equ: positivity of F}
    \end{align}    
\end{thm}

Now, we prove the direct part.
\begin{IEEEproof}[Proof of the direct part]
    We show that any pair $(r_1, r_2)$ satisfying
    \begin{align}
        (r_1, r_2) \in \left\{(R_1,R_2): R_i \geq \sup_{P_{XY}\in\cS} \left(H_{i}(X, Y) + \Delta_{\bfd} I(X;Y)\right),\ \forall i \in \{1, 2, 3\} \right\}
        \label{equ:rate_pair_in_direct_region}
    \end{align}
    is achievable.
    
    For an arbitrarily fixed $\gamma > 0$ and $(r_1, r_2)$ satisfying \eqref{equ:rate_pair_in_direct_region}, let
    \begin{align*}
        R_i = r_i + \gamma,\ \forall i \in \{1, 2\}.
    \end{align*}
    Then, according to Theorem \ref{thm:existence of seq of mixed codes}, there exists a sequence of codes $\{(f_n^{(1)}, f_n^{(2)}, \varphi_n)\}$ such that
    \begin{align*}
        M_n^{(1)} = \left\lceil 2^{n R_1}\right\rceil,\
        M_n^{(2)} = \left\lceil 2^{n R_2}\right\rceil,\
        \forall n>0,
    \end{align*}
    and
    \begin{align*}
        \liminf_{n\ra\infty}-\frac{1}{n} \log \sup_{P_{X Y} \in \cS} \max_{d \in \cD_{n}} \vep_{XY_{(d)}}^{(n)}(f_n^{(1)}, f_n^{(2)}, \varphi_n)
        &\geq \sup_{\delta > 0} \min_{i\in\{1,2,3\}} \sup_{0\leq \rho \leq 1}\inf_{P_{XY}\in\cS_{\delta}} F_{i}(\rho, R_i, P_{XY}, \Delta_{\bfd}).
    \end{align*}
    
    On the other hand, by noticing that
    \begin{align*}
        R_i &\geq \sup_{P_{XY}\in\cS} \left(H_{i}(X, Y)+\Delta_{\bfd} I(X;Y)\right) + \gamma,\ \forall i \in \{1, 2, 3\},
    \end{align*}
    Theorem \ref{thm: positivity of E} tells us that
    \begin{align*}
        \sup_{\delta > 0} \min_{i\in\{1,2,3\}} \sup_{0\leq \rho \leq 1}\inf_{P_{XY}\in\cS_{\delta}} F_{i}(\rho, R_i, P_{XY}, \Delta_{\bfd}) > 0.
    \end{align*}
    Thus, we have
    \begin{align*}
        \liminf_{n\ra\infty} -\frac{1}{n} \log \sup_{P_{X Y} \in \cS} \max_{d \in \cD_{n}} \vep_{XY_{(d)}}^{(n)}(f_n^{(1)}, f_n^{(2)}, \varphi_n) > 0.
    \end{align*}
    This impels that there exists an integer $N \geq 1$ such that for any $n \geq N$,
    \begin{align*}
        \sup_{P_{XY} \in\cS} \max_{d \in\cD_{n}} \vep_{XY_{(d)}}^{(n)}(f_n^{(1)}, f_n^{(2)}, \varphi_n) \leq \gamma.
    \end{align*}
    
    Now, we choose a sequence $\{\gamma_k\}_{k = 1}^\infty$ such that $\gamma_1 > \gamma_2 > \cdots > 0$ and $\gamma_{k} \to 0$ as $k \to \infty$. Then, by repeating the above argument for each $k \geq 1$, we can show the existence of a sequence of codes $\{(f_{n, k}^{(1)}, f_{n, k}^{(2)}, \varphi_{n, k})\}$ and an integer $N_{k} \geq 1$ such that for any $n \geq N_{k}$,
    \begin{align*}
        M_n^{(i)} &= \left\lceil 2^{n (r_i + \gamma_{k})}\right\rceil,\ \forall i \in \{1, 2\},\\
        \sup_{P_{XY} \in\cS} \max_{d \in\cD_{n}} \vep_{XY_{(d)}}^{(n)}(f_{n, k}^{(1)}, f_{n, k}^{(2)}, \varphi_{n, k}) &\leq \gamma_{k}.
    \end{align*}
    Without loss of generality, we assume that $N_{1} < N_{2} < \cdots \to \infty$. We denote by $k_{n}$ the integer $k$ satisfying $N_{k} \leq n < N_{k + 1}$ and define a code as
    $(f_{n}^{(1)}, f_{n}^{(2)}, \varphi_{n}) \teq (f_{n, k_{n}}^{(1)}, f_{n, k_{n}}^{(2)}, \varphi_{n, k_{n}})$. Then, for the sequence of these codes $\{(f_{n}^{(1)}, f_{n}^{(2)}, \varphi_{n})\}$ and any $n \geq N_1$, we have
    \begin{align*}
        M_n^{(i)} &= \left\lceil 2^{n (r_i + \gamma_{k_{n}})}\right\rceil,\ \forall i \in \{1, 2\},\\
        \sup_{P_{XY} \in\cS} \max_{d \in\cD_{n}} \vep_{XY_{(d)}}^{(n)}(f_n^{(1)}, f_n^{(2)}, \varphi_n) &\leq \gamma_{k_{n}}.
    \end{align*}
    Since $\gamma_{k_{n}} \to 0$ as $n \to \infty$, we have
    \begin{align*}
        \limsup_{n\ra\infty } R_n^{(i)} &= r_{i},\ \forall i \in \{1, 2\},\\
        \lim_{n\ra\infty} \sup_{P_{XY} \in\cS} \max_{d \in\cD_{n}}
        \vep_{XY_{(d)}}^{(n)}(f_n^{(1)}, f_n^{(2)}, \varphi_n) &= 0.
    \end{align*}
    This means that $(r_1, r_2)$ is achievable and completes the proof.
\end{IEEEproof}

\section{Elimination of Knowledge of the Bound of Delays}
\label{sec:elimination_knowledge}
In this section, we give an extension of our coding scheme which does not require knowledge of the bound of delays if possible delays are bounded. We note that this scheme is inspired by \cite{cover1981asynchronous}. 

We assume that delays are bounded, i.e., for the bound $\bfd = \{(\underline d_{n}, \overline d_{n})\}_{n=1}^\infty$, there are some constants $\overline{d}$ and $\underline{d}$ such that
\begin{align}
    \underline{d}\leq \underline{d}_{n} \leq \overline{d}_{n} \leq \overline{d},\ \forall n>0.
    \label{equ:delays_are_bounded}
\end{align}
In this case, according to Theorem \ref{thm:existence of seq of mixed codes}, there exists a sequence $\{(f_n^{(1)}, f_n^{(2)}, \varphi_n)\}$ of codes such that the error exponent is bounded as
\begin{align*}
    \liminf_{n\ra\infty}-\frac{1}{n} \log \sup_{P_{XY}\in\cS}\max_{d\in\cD_n}
    \vep_{XY_{(d)}}^{(n)}(f_n^{(1)}, f_n^{(2)}, \varphi_n)
    \geq \sup_{\delta > 0} \min_{i\in\{1,2,3\}} \sup_{0\leq \rho \leq 1}\inf_{P_{XY}\in\cS_{\delta}} F_{i}(\rho, R_i, P_{XY}, 0).
\end{align*}
We can show the existence of a sequence of codes achieving the same bound even if there is no knowledge of the bound $\bfd$.

To this end, we introduce the \textit{dummy} bound $\widetilde \bfd = \{(\underline{b}_n, \overline{b}_n)\}$ of delays, where $\underline{b}_n = - \lceil \sqrt{n} \rceil$ and $\overline{b}_n = \lceil \sqrt{n} \rceil$. For this bound $\widetilde \bfd$, according to Theorem \ref{thm:existence of seq of mixed codes}, there exists a sequence $\{(\widetilde f_n^{(1)}, \widetilde f_n^{(2)},\widetilde \varphi_n)\}$ of codes such that the error exponent is bounded as, by the same exponent,
\begin{align*}
    \liminf_{n\ra\infty}-\frac{1}{n} \log \sup_{P_{XY}\in\cS}\max_{d\in\widetilde \cD_n}
    \vep_{XY_{(d)}}^{(n)}(\widetilde f_n^{(1)},\widetilde f_n^{(2)},\widetilde \varphi_n)
    \geq \sup_{\delta > 0} \min_{i\in\{1,2,3\}} \sup_{0\leq \rho \leq 1}\inf_{P_{XY}\in\cS_{\delta}} F_{i}(\rho, R_i, P_{XY}, 0),
\end{align*}
where
\begin{align*}
    \widetilde \cD_n \teq \left\{\underline{b}_n, \underline{b}_n + 1, \cdots, \overline{b}_n - 1, \overline{b}_n \right\}.
\end{align*}
Since for sufficiently large $n$, dummy bounds $\overline{b}_n$ and $\underline{b}_n$
exceed $\overline{d}$ and $\underline{d}$, i.e., $\cD_{n}\subseteq \widetilde \cD_n$, we have
\begin{align*}
    \liminf_{n\ra\infty}-\frac{1}{n} \log \sup_{P_{XY}\in\cS}\max_{d\in\cD_n} \vep_{XY_{(d)}}^{(n)}(\widetilde f_n^{(1)},\widetilde f_n^{(2)},\widetilde \varphi_n)
    \geq  \liminf_{n\ra\infty}-\frac{1}{n} \log \sup_{P_{XY}\in\cS}\max_{d \in \widetilde \cD_n} \vep_{XY_{(d)}}^{(n)}(\widetilde f_n^{(1)},\widetilde f_n^{(2)},\widetilde \varphi_n).
\end{align*}
We note that this inequality holds for any bound $\bfd$ satisfying \eqref{equ:delays_are_bounded}, and the sequence of codes $\{(\widetilde f_n^{(1)},\widetilde f_n^{(2)},\widetilde \varphi_n)\}$ depends on the dummy bound $\widetilde \bfd$ and is independent of the true bound $\bfd$. Thus, we have the following theorem.

\begin{thm}\rm
    For any $R_1, R_2>0$, there exists a sequence $\{(\widetilde f_n^{(1)},\widetilde f_n^{(2)},\widetilde \varphi_n)\}$ of codes such that
    \begin{align*}
        M_n^{(1)} = \left\lceil 2^{n R_1}\right\rceil,\
        M_n^{(2)} = \left\lceil 2^{n R_2}\right\rceil,\
        \forall n>0,
    \end{align*}
    and
    \begin{align*}
        \liminf_{n\ra\infty}-\frac{1}{n} \log \sup_{P_{XY}\in\cS}\max_{d\in\cD_n} \vep_{XY_{(d)}}^{(n)}(\widetilde f_n^{(1)},\widetilde f_n^{(2)},\widetilde \varphi_n)
        \geq \sup_{\delta > 0} \min_{i\in\{1,2,3\}} \sup_{0\leq \rho \leq 1}\inf_{P_{XY}\in\cS_{\delta}} F_{i}(\rho, R_i, P_{XY}, 0),
    \end{align*}
    where the sequence $\{(\widetilde f_n^{(1)},\widetilde f_n^{(2)},\widetilde \varphi_n)\}$ is independent of $\bfd$, and the inequality holds for any bound $\bfd = \{(\underline d_{n}, \overline d_{n})\}_{n=1}^\infty$ such that there are some constants $\overline{d}$ and $\underline{d}$ satisfying
    \begin{align*}
        \underline{d}\leq \underline{d}_{n} \leq \overline{d}_{n} \leq \overline{d},\ \forall n>0.
    \end{align*}    
\end{thm}

\section{Conclusion}
\label{sec:conclusion}
In this paper, we have dealt with the asynchronous SW coding system and clarified the achievable rate region. According to the achievable rate region, it does not always coincide with that of the synchronous SW coding system. In the converse part, we used a usual information-theoretic technique. In the direct part, we employed a universal coding scheme using the mixed source. We extended our coding scheme to a scheme which does not require knowledge of the bound of delays.

\appendices
\section{Converse Bound} \label{sec:converse_bound}
In this appendix, we prove the bound \eqref{equ:lower_from_Fano's_ineq} in a similar way as in \cite[Sec.\ 15.4.2]{cover2006eit}.

Due to Fano's inequality \cite[Theorem 2.10.1]{cover2006eit}, for any $n > 0$, code $(f_n^{(1)}, f_n^{(2)}, \varphi_n)$, delay $d\in\cD_{n}$, and $P_{XY}\in\cS$, we have
\begin{align}
    H(X^n,Y^n_{(d)}|f_n^{(1)}(X^{n}),f_n^{(2)}(Y_{(d)}^{n}))
    &\leq n \vep_{XY_{(d)}}^{(n)} \log |\cX||\cY| + 1.
    \label{equ:app:conv3}
\end{align}
We also have
\begin{align}
    H(X^n|f_n^{(1)}(X^{n}),Y^n_{(d)})
    &= H(X^n|f_n^{(1)}(X^{n}),f_n^{(2)}(Y_{(d)}^{n}),Y^n_{(d)})\notag\\
    &\leq H(X^n,Y^n_{(d)}|f_n^{(1)}(X^{n}),f_n^{(2)}(Y_{(d)}^{n}))\notag\\
    &\leq n \vep_{XY_{(d)}}^{(n)} \log |\cX||\cY| + 1,
    \label{equ:app:conv1}
\end{align}
and similarly
\begin{align}
    H(Y^n_{(d)}|f_n^{(2)}(Y^n_{(d)}),X^n)
    &\leq n \vep_{XY_{(d)}}^{(n)} \log |\cX||\cY| + 1.
    \label{equ:app:conv2}
\end{align}
Thus, we have
\begin{align}
    nR_n^{(1)}
    &\geq H(f_n^{(1)}(X^{n}))\notag\\
    &\geq H(f_n^{(1)}(X^{n})|Y^n_{(d)})\notag\\
    &= I(X^n;f_n^{(1)}(X^{n})|Y^n_{(d)})\notag\\
    &=H(X^n|Y^n_{(d)}) - H(X^n|f_n^{(1)}(X^{n}),Y^n_{(d)})\notag\\
    &\geq H(X^n|Y^n_{(d)}) -  n \vep_{XY_{(d)}}^{(n)} \log |\cX||\cY| - 1,
    \label{equ:app:conv_bound1}
\end{align}
where the last inequality comes from \eqref{equ:app:conv1}. Similarly, due to \eqref{equ:app:conv2}, we have
\begin{align}
    nR_n^{(2)}
    &\geq H(Y^n_{(d)}|X^n) -  n \vep_{XY_{(d)}}^{(n)} \log |\cX||\cY| - 1.
    \label{equ:app:conv_bound2}
\end{align}
We also have
\begin{align}
    n (R_n^{(1)} + R_n^{(2)})
    &\geq H(f_n^{(1)}(X^{n}),f_n^{(2)}(Y^{n}_{(d)}))\notag\\
    &= I(X^n,Y^n_{(d)};f_n^{(1)}(X^{n}), f_n^{(2)}(Y^{n}_{(d)}))\notag\\
    &= H(X^n, Y^n_{(d)}) - H(X^n,Y^n_{(d)} |f_n^{(1)}(X^{n}),f_n^{(2)}(Y^{n}_{(d)}))\notag\\
    &\geq H(X^n, Y^n_{(d)}) - n \vep_{XY_{(d)}}^{(n)} \log |\cX||\cY| - 1,
    \label{equ:app:conv_bound3}
\end{align}
where the last inequality comes from \eqref{equ:app:conv3}.

Combining \eqref{equ:app:conv_bound1}--\eqref{equ:app:conv_bound3}, we have \eqref{equ:lower_from_Fano's_ineq}.
    
\section{Proof of Lemma \ref{lem: exists of type for delay}}
\label{sec:proof of lem: exists of type for delay}
We first give the next lemma which is an extension of \cite[Lemma 3]{matsuta2010universalIEICE} to non-binary alphabets.
\begin{lem}
    \rm\label{lem: exists of type}
    Let $\alpha = |\cX| (|\cX| - 1)$ and $\vep_n = \A \ln \big(1 + \frac{\A}{n - \A}\big)$. Then, for any $n\geq \alpha$ and any $P_{X} \in \cP(\cX)$,  there exists an $n$-type $P_{\bar{X}}\in \cP_{n}(\cX)$ such that
    \begin{align*}
        d_{\rm v}(P_{X}, P_{\bar{X}}) \leq \frac{2 (|\cX| - 1)}{n},
    \end{align*}
    and
    \begin{align}
        P_{X}^{n}(x^n) \leq P_{\bar X}^{n}(x^n)e^{(\A + \vep_n)},\ \forall x^n \in \cX^n.
        \label{eqn: n-fold bound by the type}
    \end{align}
\end{lem}
\begin{IEEEproof}
    For any fixed $P_{X}\in \cP(\cX)$, let $x^*\in\cX$ be a symbol with the highest probability, i.e. $P_{X}(x^*)\geq P_{X}(x)$ for any $x\in \cX$. Let $N(x)$ be an integer defined as
    \begin{align}
        N(x)\teq 
        \begin{cases}
            \left \lceil n P_{X}(x) \right \rceil & \mbox{ if } x \neq x^*,\\
            n - \sum_{x\in\cX: x\neq x^*} \left \lceil n P_{X}(x) \right \rceil &\mbox{ if } x = x^*.
        \end{cases}
        \label{equ: def: N(x)}
    \end{align}
    Then, we have
    \begin{align*}
        0 \leq N(x) \leq n,\ \forall x \neq x^*,
    \end{align*}
    and
    \begin{align*}
        n &\geq N(x^*) \\
        &= n - \sum_{x\in\cX: x\neq x^*} \left \lceil n P_{X}(x) \right \rceil\\
        &\geq n - \sum_{x\in\cX: x\neq x^*} (n P_{X}(x) +1 )\\
        &=  n \left(1 - \sum_{x\in\cX: x\neq x^*} P_{X}(x)\right)  - (|\cX|-1)\\
        &=  n P_{X}(x^*)  - (|\cX|-1)\\
        &\geq  n \frac{1}{|\cX|}  - (|\cX|-1)\\
        &=  \frac{n-|\cX|(|\cX|-1)}{|\cX|}.
    \end{align*}
    Thus, for any $n\geq |\cX|(|\cX|-1) = \alpha$, we have
    \begin{align*}
        0 \leq N(x) \leq n,\ \forall x\in\cX,
    \end{align*}
    and also
    \begin{align*}
        \sum_{x\in\cX} N(x) = n.
    \end{align*}

    For this $N(x)$, we define an $n$-type $P_{\bar X}\in\cP_n(\cX)$ as
    \begin{align*}
        P_{\bar X}(x) &\teq \frac{N(x)}{n}.
    \end{align*}
    Then, according to (\ref{equ: def: N(x)}), we have
    \begin{align}
        0 \leq P_{\bar{X}}(x)-P_{X}(x) < \frac{1}{n},\ \forall x \neq x^*, \label{equ:compPP2}
    \end{align}
    and
    \begin{align}
        0 \leq P_{X}(x^*) - P_{\bar{X}}(x^*)
        < \frac{|\cX|-1}{n}.\label{equ:distancePPvar}
    \end{align}
    Hence, we have
    \begin{align*}
        d_{\rm v}(P_{X}, P_{\bar{X}}) = \sum_{x\in\cX} |P_{\bar{X}}(x)-P_{X}(x)| < \frac{2(|\cX|-1)}{n}.
    \end{align*}
    
    We also show that the $n$-type $P_{\bar{X}}$ satisfies (\ref{eqn: n-fold bound by the type}).
    For any $x^{n}\in\cX^n$ and $x \in \cX$,
    define $N(x|x^{n})$ as 
    \begin{align*}
        N(x|x^{n}) \teq |\{i \in \{1, \cdots, n\}: x_i = x\}|.
    \end{align*}
    Then, for any $x^{n}\in\cX^n$, we have
    \begin{align*}
        P^n_{X}(x^{n})
        & = \prod_{x\in\cX} P_{X}(x)^{N(x|x^{n})}\\
        & = P_{X}(x^*)^{N(x^*|x^{n})} \prod_{x\in\cX: x \neq x^*} P_{X}(x)^{N(x|x^{n})}\\
        & \leq \left(P_{\bar{X}}(x^*) + \frac{|\cX|-1}{n} \right)^{N(x^*|x^{n})} \prod_{x\in\cX: x \neq x^*} P_{\bar{X}}(x)^{N(x|x^{n})}\\
        & =  P_{\bar{X}}(x^*)^{N(x^*|x^{n})} \left(1 + \frac{|\cX|-1}{n P_{\bar{X}}(x^*)}  \right)^{N(x^*|x^{n})} \prod_{x\in\cX: x \neq x^*} P_{\bar{X}}(x)^{N(x|x^{n})}\\
        & = P^n_{\bar{X}}(x^{n})
        \left(1 + \frac{|\cX|-1}{n P_{\bar{X}}(x^*)} \right)^{N(x^*|x^{n})},
    \end{align*}
    where the inequality comes from \eqref{equ:compPP2} and \eqref{equ:distancePPvar}. According to (\ref{equ:distancePPvar}) and the fact that $P_{X}(x^*)\geq \frac{1}{|\cX|}$, we have
    \begin{align*}
        \frac{|\cX|-1}{n P_{\bar{X}}(x^*)}
        &<  \frac{|\cX|-1}{n P_{X}(x^*)-(|\cX|-1)}\\
        &\leq  \frac{\A}{n-\A}.
    \end{align*}
    Since $N(x^*|x^{n})\leq n$, we have
    \begin{align*}
        P^n_{X}(x^{n})
        &\leq P^n_{\bar{X}}(x^{n})
        \bigg(1 + \frac{|\cX|-1}{n P_{\bar{X}}(x^*)}
        \bigg)^{N(x^*|x^{n})}\\
        &\leq P^n_{\bar{X}}(x^{n})
        \bigg(1 + \frac{{\A}}{n-{\A}} \bigg)^{N(x^*|x^{n})}\\
        &\leq P^n_{\bar{X}}(x^{n})
        \bigg(1 + \frac{\A}{n-\A} \bigg)^n\\
        &= P^n_{\bar{X}}(x^{n})
        \bigg(1 + \frac{1}{n'} \bigg)^{\A n'+\A}\\
        &= P^n_{\bar{X}}(x^{n})
        \bigg(1 + \frac{\A}{n-\A} \bigg)^{\A}\bigg(1 + \frac{1}{n'} \bigg)^{\A n'},
    \end{align*}
    where $n' = n/\A-1$. Since $\Big(1 + \frac{1}{n'} \Big)^{n'}$ is monotonically increasing function of $n'$ and converges to $e$, we have
    \begin{align*}
        P^n_{X}(x^{n})
        &\leq P^n_{\bar{X}}(x^{n})
        \Big(1 + \frac{\A}{n-\A} \Big)^{\A} e^{\A}\\
        &= P^n_{\bar{X}}(x^{n}) e^{\A+\vep_n}.
    \end{align*}
    This completes the proof.
\end{IEEEproof}

By using the above lemma, we prove Lemma \ref{lem: exists of type for delay}.
\begin{IEEEproof}[Proof of Lemma \ref{lem: exists of type for delay}]
    According to Lemma \ref{lem: exists of type}, for any $n \geq \alpha\ (= |\cX| |\cY| (|\cX| |\cY| - 1))$, there exists an $n$-type $P_{\bar{X}\bar{Y}}\in \cP_n(\cX\times\cY)$ such that
    \begin{align}
        P^n_{XY}({x^n},{y^n})
        &\leq  P^n_{\bar{X}\bar{Y}}({x^n},{y^n}) e^{\A+\vep_n},
        \label{equ:distPPvarnodelay}
    \end{align}
    and
    \begin{align*}
        d_{\rm v}(P_{XY}, P_{\bar{X}\bar{Y}}) \leq \frac{2(|\cX| |\cY|-1)}{n}.
    \end{align*}
    Note that, for any $k\leq n$, marginal distributions $P^k_{XY}({x^k},{y^k})$, $P^k_{X}({x^k})$, and $P^k_{Y}({y^k})$ of $P^n_{XY}({x^n},{y^n})$ satisfy
    \begin{align*}
        P^k_{XY}({x^k},{y^k})
        &\leq  P^k_{\bar{X}\bar{Y}}({x^k},{y^k}) e^{\A+\vep_n}, \\
        P^k_{X}({x^k})
        &\leq  P^k_{\bar{X}}({x^k}) e^{\A+\vep_n}, \\
        P^k_{Y}({y^k})
        &\leq  P^k_{\bar{Y}}({y^k}) e^{\A+\vep_n}.
    \end{align*}
    Thus, for any $d \in \{-n, -n + 1, \cdots, n\}$, we have
    \begin{align*}
        P_{X^{n}Y_{(d)}^{n}}(x^n,y^n)
        &= P_{X}^{|d|}(x_{\rm I}^{|d|}) P_{XY}^{n-|d|}(x_{\rm C}^{n - |d|}, y_{\rm C}^{n-|d|}) P_{Y}^{|d|}(y_{\rm I}^{|d|})\\
        &\leq P_{\bar X}^{|d|}(x_{\rm I}^{|d|}) P_{\bar X \bar Y}^{n-|d|}(x_{\rm C}^{n - |d|}, y_{\rm C}^{n-|d|}) P_{\bar Y}^{|d|}(y_{\rm I}^{|d|}) e^{3(\A+\vep_n)}\\
        &= P_{\bar X^{n}\bar Y_{(d)}^{n}}(x^n,y^n) e^{3(\A+\vep_n)}.
    \end{align*}
    This completes the proof.
\end{IEEEproof}

\section{Gallager-Type Bounds}
\label{appendix: GTBMS}
In this appendix, we show the bound \eqref{eq:ExBound1} in the proof of Theorem \ref{thm:existence of mixed code}. First of all, we give the next two lemmas.
\begin{lem}
    \label{lem: bound of E_n}
    For any finite set $\cS_n\subseteq \cP(\cX^n\times\cY^n)$ of PMFs and the mixture $Q_n$ of all PMFs in $\cS_n$, i.e.,
    \begin{align*}
        Q_n(x^n,y^n) = \sum_{P_n\in\cS_n} \frac{1}{|\cS_n|}P_n(x^n,y^n),
    \end{align*} 
    we have for any $\rho \in [0, 1]$,
    \begin{align}
        E_n^{(i)}(\rho, Q_{n}) \leq \frac{1}{n} \log |\cS_n| + \max_{P_n \in  \cS_n} E_n^{(i)}(\rho, P_n),\ \forall i \in \{1, 2, 3\}.
        \label{app:GTBMS:equ:mixed_exponent}
    \end{align}
\end{lem}
\begin{IEEEproof}
    For any $\rho \in [0, 1]$, we have
    \begin{align}
        E_n^{(1)}(\rho,Q_{n})
        & = \frac{1}{n}\log 
        \sum_{y^n\in\cY^n} \left(\sum_{x^n\in\cX^n}Q_{n}(x^n,y^n)^{\frac{1}{1 + \rho}}\right)^{1 + \rho}\notag\\
        & = \frac{1}{n}\log \sum_{y^n\in\cY^n} \left(\sum_{x^n\in\cX^n}\left(\sum_{P_{n} \in \cS_n} \frac{1}{|\cS_n|}
                P_n(x^n,y^n)\right)^{\frac{1}{1 + \rho}}\right)^{1 + \rho}\notag\\
        & = \frac{1}{n}\log \frac{1}{|\cS_n|}
        \sum_{y^n\in\cY^n} \left(\sum_{x^n\in\cX^n}\left(\sum_{P_{n} \in \cS_n} 
                P_n(x^n,y^n)\right)^{\frac{1}{1 + \rho}}\right)^{1 + \rho}\notag\\
        &\overset{\rm (a)}{\leq} \frac{1}{n}\log \frac{1}{|\cS_n|}
        \sum_{y^n\in\cY^n} \left(\sum_{x^n\in\cX^n}
            \sum_{P_{n} \in \cS_n} 
            P_n(x^n,y^n)^{\frac{1}{1 + \rho}}\right)^{1 + \rho}\notag\\
        & \overset{\rm (b)}{\leq}
        \frac{1}{n}\log \frac{1}{|\cS_n|}
        \sum_{y^n\in\cY^n}
        \sum_{P_{n} \in \cS_n} 
        \frac{1}{|\cS_n|} 
        \left(|\cS_n|
            \sum_{x^n\in\cX^n} 
            P_n(x^n,y^n)^{\frac{1}{1 + \rho}}\right)^{1 + \rho}\notag\\
        & = \frac{1}{n}\log \frac{|\cS_n|^{1 + \rho}}{|\cS_n|^2}
        \sum_{P_{n} \in \cS_n} 
        \sum_{y^n\in\cY^n}
        \left( \sum_{x^n\in\cX^n} 
            P_n(x^n,y^n)^{\frac{1}{1 + \rho}}\right)^{1 + \rho}\notag\\
        & \leq \frac{1}{n}\log \frac{|\cS_n|^{1 + \rho}}{|\cS_n|^2}
        |\cS_n| \max_{P_{n} \in \cS_n}
        \sum_{y^n\in\cY^n}
        \left( \sum_{x^n\in\cX^n} 
            P_n(x^n,y^n)^{\frac{1}{1 + \rho}}\right)^{1 + \rho}\notag\\
        & = \frac{1}{n}\log  |\cS_n|^{\rho} \max_{P_{n} \in \cS_n}
        \sum_{y^n\in\cY^n}
        \left( \sum_{x^n\in\cX^n} 
            P_n(x^n,y^n)^{\frac{1}{1 + \rho}}\right)^{1 + \rho}\notag\\
        & \leq \frac{1}{n}\log  |\cS_n| \max_{P_{n} \in \cS_n}
        \sum_{y^n\in\cY^n}
        \left( \sum_{x^n\in\cX^n} 
            P_n(x^n,y^n)^{\frac{1}{1 + \rho}}\right)^{1 + \rho}\notag\\
        & = \frac{1}{n}\log  |\cS_n| +\max_{P_{n} \in \cS_n}E_n^{(1)}(\rho,P_n),
        \label{app:GTBMS:equ:mixed_exponent1}
    \end{align}
    where (a) follows since $(a+b)^r \leq a^r+b^r$ for $0\leq r\leq 1$ and $a,b\geq 0$ (see Lemma \ref{lem: hardy's inequality} in Appendix \ref{appendix: uniform continuity}), and (b) comes from Jensen's inequality (see, e.g., \cite{cover2006eit}). Similarly, we have
    \begin{align}
        E_n^{(2)}(\rho,Q_{n})
        \leq \frac{1}{n}\log  |\cS_n| +\max_{P_n\in \cS_n}E_n^{(2)}(\rho,P_n).
        \label{app:GTBMS:equ:mixed_exponent2}
    \end{align}

    We also have
    \begin{align}
        E_n^{(3)}(\rho, Q_{n})
        & = \frac{1}{n}\log  \left(\sum_{(x^n, y^n) \in \cX^n \times \cY^n}Q_{n}(x^n,y^n)^{\frac{1}{1 + \rho}}\right)^{1 + \rho}\notag\\
        & = \frac{1}{n}\log  \left(\sum_{(x^n, y^n) \in \cX^n \times \cY^n}\left(\sum_{P_{n} \in \cS_n} \frac{1}{|\cS_n|} P_n(x^n,y^n)\right)^{\frac{1}{1 + \rho}}\right)^{1 + \rho}\notag\\
        & = \frac{1}{n}\log \frac{1}{|\cS_n|} \left(\sum_{(x^n, y^n) \in \cX^n \times \cY^n}\left(\sum_{P_{n} \in \cS_n} P_n(x^n,y^n)\right)^{\frac{1}{1 + \rho}}\right)^{1 + \rho}\notag\\
        &\overset{\rm (a)}{\leq} \frac{1}{n}\log \frac{1}{|\cS_n|} \left(\sum_{(x^n, y^n) \in \cX^n \times \cY^n} \sum_{P_{n} \in \cS_n} P_n(x^n,y^n)^{\frac{1}{1 + \rho}}\right)^{1 + \rho}\notag\\
        & \overset{\rm (b)}{\leq} \frac{1}{n}\log \frac{1}{|\cS_n|} \sum_{P_{n} \in \cS_n} \frac{1}{|\cS_n|} \left(|\cS_n|\sum_{(x^n, y^n) \in \cX^n \times \cY^n} P_n(x^n,y^n)^{\frac{1}{1 + \rho}}\right)^{1 + \rho}\notag\\
        & = \frac{1}{n}\log \frac{|\cS_n|^{1 + \rho}}{|\cS_n|^2} \sum_{P_{n} \in \cS_n} \left( \sum_{(x^n, y^n) \in \cX^n \times \cY^n} P_n(x^n,y^n)^{\frac{1}{1 + \rho}}\right)^{1 + \rho}\notag\\
        & \leq \frac{1}{n}\log \frac{|\cS_n|^{1 + \rho}}{|\cS_n|^2} |\cS_n| \max_{P_{n} \in \cS_n} \left( \sum_{(x^n, y^n) \in \cX^n \times \cY^n} P_n(x^n,y^n)^{\frac{1}{1 + \rho}}\right)^{1 + \rho}\notag\\
        & = \frac{1}{n}\log  |\cS_n|^{\rho} \max_{P_{n} \in \cS_n} \left( \sum_{(x^n, y^n) \in \cX^n \times \cY^n} P_n(x^n,y^n)^{\frac{1}{1 + \rho}}\right)^{1 + \rho}\notag\\
        & \leq \frac{1}{n}\log  |\cS_n| \max_{P_{n} \in \cS_n} \left( \sum_{(x^n, y^n) \in \cX^n \times \cY^n} P_n(x^n,y^n)^{\frac{1}{1 + \rho}}\right)^{1 + \rho}\notag\\
        & = \frac{1}{n} \log  |\cS_n| +\max_{P_n\in \cS_n}E_n^{(3)}(\rho,P_n),
        \label{app:GTBMS:equ:mixed_exponent3}
    \end{align}
    where (a) and (b) come from the same reason as above.

    Combining \eqref{app:GTBMS:equ:mixed_exponent1}--\eqref{app:GTBMS:equ:mixed_exponent3}, we have \eqref{app:GTBMS:equ:mixed_exponent}.
\end{IEEEproof}

The next lemma gives identities by Gallager (cf.\ \cite[Section II]{gallager1976}).
\begin{lem}
    \label{lem: H - E = D}
    For any $P_{XY}\in\cP(\cX\times\cY)$ and $\rho \in [0, 1]$, it holds that
    \begin{align}
        \rho H(P_{X}^{(\rho)}) - E^{(3)}(\rho,P_{X}) &= D(P_{X}^{(\rho)}\|P_{X}), \label{equ:lem: H - E = D 1}\\
        \rho H(P_{X|Y}^{(\rho)} | \bar P_{Y}^{(\rho)}) - E^{(1)}(\rho,P_{XY}) &= D(\bar P_{Y}^{(\rho)}\times P_{X|Y}^{(\rho)}\|P_{XY}), \label{equ:lem: H - E = D 2}\\
        \rho H(P_{Y|X}^{(\rho)} | \bar P_{X}^{(\rho)}) - E^{(2)}(\rho,P_{XY}) &= D(\bar P_{X}^{(\rho)}\times P_{Y|X}^{(\rho)}\|P_{XY}), \label{equ:lem: H - E = D 3}
    \end{align}
    where
    \begin{align*}
        E^{(3)}(\rho,P_{X}) &\teq E_{1}^{(3)}(\rho,P_{X}) = \log \left( \sum_{x\in\cX}  P_{X}(x)^{\frac{1}{1 + \rho}}\right)^{1 + \rho},\\
        E^{(1)}(\rho,P_{XY}) &\teq E_{1}^{(1)}(\rho,P_{XY}) = \log \sum_{y\in\cY}
        \left(\sum_{x\in\cX} P_{XY}(x, y)^{\frac{1}{1 + \rho}}\right)^{1 + \rho},\\
        E^{(2)}(\rho,P_{XY}) &\teq E_{1}^{(2)}(\rho,P_{XY}) = \log \sum_{x\in\cX}
        \left(\sum_{y\in\cY} P_{XY}(x, y)^{\frac{1}{1 + \rho}}\right)^{1 + \rho}.
    \end{align*}
\end{lem}
\begin{IEEEproof}
    Since $P_{X}^{(\rho)}(x)=\frac{P_{X}(x)^{\frac{1}{1+\rho}}}{\sum_{x\in\cX}P_{X}(x)^{\frac{1}{1+\rho}}}$,
    we have for any $x\in\cX$ such that $P_{X}(x) > 0$,
    \begin{align*}
        \left( \sum_{x\in\cX} P_{X}(x)^{\frac{1}{1+\rho}}\right)^{1+\rho}
        &= \left( \frac{P_{X}(x)^{\frac{1}{1+\rho}}}{P_{X}^{(\rho)}(x)}\right)^{1+\rho}\notag\\
        &= \frac{P_{X}(x)}{P_{X}^{(\rho)}(x)^{1+\rho}}.
    \end{align*}
    Thus, we have
    \begin{align*}
        - \log \left( \sum_{x\in\cX} P_{X}(x)^{\frac{1}{1+\rho}}\right)^{1+\rho}
        &= - \sum_{x\in\cX: P_{X}(x) > 0} P_{X}^{(\rho)}(x) \log \frac{P_{X}(x)}{P_{X}^{(\rho)}(x)^{1+\rho}}\notag\\
        & = \sum_{x\in\cX: P_{X}^{(\rho)}(x) > 0} P_{X}^{(\rho)}(x) \log \frac{P_{X}^{(\rho)}(x)P_{X}^{(\rho)}(x)^{\rho}}{P_{X}(x)}\notag\\
        & = D(P_{X}^{(\rho)}\|P_{X}) - \rho H(P_{X}^{(\rho)}).
    \end{align*}
    This identity gives \eqref{equ:lem: H - E = D 1}.

    On the other hand, since
    \begin{align*}
        P_{X|Y}^{(\rho)}(x|y)
        = \frac{P_{XY}(x,y)^{\frac{1}{1+\rho}}}
        {\sum_{x\in\cX} P_{XY}(x,y)^{\frac{1}{1+\rho}}}
    \end{align*}
    and
    \begin{align*}
        \bar P_{Y}^{(\rho)}(y)
        = \frac{\left(\sum_{x\in\cX} P_{XY}(x,y)^{\frac{1}{1+\rho}}\right)^{1+\rho}}
        {\sum_{y\in\cY}\left(\sum_{x\in\cX} P_{XY}(x,y)^{\frac{1}{1+\rho}}\right)^{1+\rho}}, 
    \end{align*}
    we have for any $(x,y)\in\cX\times\cY$ such that $P_{XY}(x,y)>0$,
    \begin{align*}
        \sum_{x\in\cX} P_{XY}(x,y)^{\frac{1}{1+\rho}}
        = \frac{P_{XY}(x,y)^{\frac{1}{1+\rho}}}{P_{X|Y}^{(\rho)}(x|y)},
    \end{align*}
    and
    \begin{align*}
        \sum_{y\in\cY}\left(\sum_{x\in\cX} P_{XY}(x,y)^{\frac{1}{1+\rho}}\right)^{1+\rho}
        &= \frac{\left(\sum_{x\in\cX} P_{XY}(x,y)^{\frac{1}{1+\rho}}\right)^{1+\rho}}{\bar P_{Y}^{(\rho)}(y)}\\
        &= \frac{\left(\frac{P_{XY}(x,y)^{\frac{1}{1+\rho}}}{P_{X|Y}^{(\rho)}(x|y)}\right)^{1+\rho}}{\bar P_{Y}^{(\rho)}(y)}\\
        &= \frac{P_{XY}(x,y)}{\bar P_{Y}^{(\rho)}(y)P_{X|Y}^{(\rho)}(x|y)^{1+\rho}}.
    \end{align*}
    Thus, we have
    \begin{align}
        &  - \log \sum_{y\in\cY}\left(\sum_{x\in\cX} P_{XY}(x,y)^{\frac{1}{1+\rho}}\right)^{1+\rho}\notag\\        
        &= \sum_{(x,y)\in\cX\times\cY:P_{XY}(x,y)>0} \bar P_{Y}^{(\rho)}(y)P_{X|Y}^{(\rho)}(x|y)
        \log \frac{\bar P_{Y}^{(\rho)}(y)P_{X|Y}^{(\rho)}(x|y)^{1+\rho}}{P_{XY}(x,y)}\notag\\
        &= D(\bar P_{Y}^{(\rho)}\times P_{X|Y}^{(\rho)}\|P_{XY}) -  \rho H(P_{X|Y}^{(\rho)}|\bar P_{Y}^{(\rho)}).
        \label{equ:for lem: H - E = D 2}
    \end{align}
    Similarly, we have
    \begin{align}
        &  - \log \sum_{x\in\cX}\left(\sum_{y\in\cY} P_{XY}(x,y)^{\frac{1}{1+\rho}}\right)^{1+\rho}\notag\\
        &= D(\bar P_{X}^{(\rho)}\times P_{Y|X}^{(\rho)}\|P_{XY}) -  \rho H(P_{Y|X}^{(\rho)}|\bar P_{X}^{(\rho)}).
        \label{equ:for lem: H - E = D 3}
    \end{align}
    Identities \eqref{equ:for lem: H - E = D 2} and \eqref{equ:for lem: H - E = D 3} give \eqref{equ:lem: H - E = D 2} and \eqref{equ:lem: H - E = D 3}, respectively.
\end{IEEEproof}
\begin{cor}
    \label{cor: upper bound of D}
    For any $P_{XY}\in\cP(\cX\times\cY)$ and $\rho \in [0, 1]$, it holds that
    \begin{align*}
        D(P_{X}^{(\rho)}\|P_{X}) &\leq \log |\cX|,\\
        D(\bar P_{Y}^{(\rho)}\times P_{X|Y}^{(\rho)}\|P_{XY}) &\leq \log |\cX|.
    \end{align*}
\end{cor}
\begin{IEEEproof}
    The corollary  immediately follows since $E^{(1)}(\rho, P_{X}) \geq 0$ and $E^{(3)}(\rho, P_{X Y}) \geq 0$.
\end{IEEEproof}

Now, we show the bound (\ref{eq:ExBound1}). For any $\rho \in [0, 1]$ and $i \in \{1, 2, 3\}$, we have
\begin{align}
    E_n^{(i)}(\rho,P_{\widetilde{X}^n\widetilde{Y}^n})
    &\overset{\rm (a)}{\leq}
    \frac{1}{n}\log |\bar{\cS}_{n}||\cD_n|
    + \max_{P_{XY}\in \bar{\cS}_{n}}\max_{d\in \cD_n}E_n^{(i)}(\rho,P_{X^{n}Y_{(d)}^{n}})\notag\\
    &\overset{\rm (b)}{\leq} \frac{\log (2 n + 1) (n + 1)^{|\cX| |\cY|}}{n} + \max_{P_{XY}\in \bar{\cS}_{n}}\max_{d \in \cD_n}E_n^{(i)}(\rho,P_{X^{n}Y_{(d)}^{n}})\notag\\
    &= \epsilon'_n + \max_{P_{XY}\in \bar{\cS}_{n}}\max_{d\in \cD_n}E_n^{(i)}(\rho,P_{X^{n}Y_{(d)}^{n}}),
    \label{inq:PartBound1}
\end{align}
where (a) comes form Lemma \ref{lem: bound of E_n}, (b) comes from the fact that $|\bar{\cS}_{n}| \leq (n + 1)^{|\cX| |\cY|}$ and $|\cD_n|\leq 2n+1$, and
\begin{align*}
    \epsilon'_n = \frac{\log (2 n + 1) (n + 1)^{|\cX| |\cY|}}{n}.
\end{align*}
Thus, we have
\begin{align}
    \sup_{0\leq\rho\leq 1} \left(\rho R_i - E_n^{(i)}(\rho, P_{\widetilde{X}^n \widetilde{Y}^n})\right)
    &\geq \sup_{0\leq\rho\leq 1} \left(\rho R_i - \max_{P_{XY}\in \bar{\cS}_{n}} \max_{d\in \cD_n} E_n^{(i)}(\rho,P_{X^{n}Y_{(d)}^{n}})\right) - \epsilon'_n\notag\\
    &= \sup_{0\leq\rho\leq 1} \min_{P_{XY}\in \bar{\cS}_{n}}
    \left(\rho R_i - \max_{d\in \cD_n}E_n^{(i)}(\rho,P_{X^{n}Y_{(d)}^{n}})\right) - \epsilon'_n.
    \label{equ: ExBound1-1}
\end{align}

First, we show the bound (\ref{eq:ExBound1}) for the case where $i = 1$. For any $\rho \in [0, 1]$, we have
\begin{align*}
    &E_n^{(1)}(\rho, P_{X^{n} Y_{(d)}^{n}})\\
    &= \frac{1}{n} \log \sum_{y^n \in \cY^n} \left(\sum_{x^n\in\cX^n}  \left(P_{X}^{|d|}(x_{\rm I}^{|d|}) P_{XY}^{n-|d|}(x_{\rm C}^{n-|d|}, y_{\rm C}^{n-|d|}) P_{Y}^{|d|}(y_{\rm I}^{|d|})\right)^{\frac{1}{1 + \rho}}\right)^{1 + \rho}\\
    &=
    \frac{1}{n}\log \sum_{y^n \in \cY^n}
    \left( \sum_{x_{\rm I}^{|d|}\in\cX^{|d|}} 
        P_{X}^{|d|}(x_{\rm I}^{|d|})^{\frac{1}{1 + \rho}}\right)^{1 + \rho}
    \left(\sum_{x_{\rm C}^{n-|d|}\in\cX^{n-|d|}}
        P_{XY}^{n-|d|}(x_{\rm C}^{n-|d|}, y_{\rm C}^{n-|d|})^{\frac{1}{1 + \rho}}\right)^{1 + \rho} \left(P_{Y}^{|d|}(y_{\rm I}^{|d|})^{\frac{1}{1 + \rho}}\right)^{1 + \rho}\\
    &=
    \frac{1}{n}\log 
    \left( \sum_{x_{\rm I}^{|d|}\in\cX^{|d|}} 
        P_{X}^{|d|}(x_{\rm I}^{|d|})^{\frac{1}{1 + \rho}}\right)^{1 + \rho}
    \sum_{y_{\rm C}^{n-|d|}\in\cY^{n-|d|}}
    \left(\sum_{x_{\rm C}^{n-|d|}\in\cX^{n-|d|}}
        P_{XY}^{n-|d|}(x_{\rm C}^{n-|d|}, y_{\rm C}^{n-|d|})^{\frac{1}{1 + \rho}}\right)^{1 + \rho}\\
    &=
    \frac{|d|}{n}\log 
    \left( \sum_{x\in\cX} 
        P_{X}(x)^{\frac{1}{1 + \rho}}\right)^{1 + \rho}
    + \frac{n-|d|}{n}\log 
    \sum_{y\in\cY}
    \left(\sum_{x\in\cX}
        P_{XY}(x, y)^{\frac{1}{1 + \rho}}\right)^{1 + \rho}\\
    &=
    \log \sum_{y\in\cY}
    \left(\sum_{x\in\cX}
        P_{XY}(x, y)^{\frac{1}{1 + \rho}}\right)^{1 + \rho}
    + \frac{|d|}{n}\left( \log
        \left( \sum_{x\in\cX} 
            P_{X}(x)^{\frac{1}{1 + \rho}}\right)^{1 + \rho}
        -\log
        \sum_{y\in\cY}
        \left(\sum_{x\in\cX}
            P_{XY}(x, y)^{\frac{1}{1 + \rho}}\right)^{1 + \rho}\right)\\
    &=
    E^{(1)}(\rho,P_{XY})
    + \frac{|d|}{n}\left( E^{(3)}(\rho,P_{X})- E^{(1)}(\rho,P_{XY})\right).
\end{align*}
On the other hand, due to (reverse) Minkowski's inequality (see \cite[Theorem 25]{hardy1952inequalities}), we have
\begin{align*}
    \log \bigg( \sum_{x \in \cX} 
    P_{X}(x)^{\frac{1}{1 + \rho}}\bigg)^{1 + \rho} \geq \log  \sum_{y \in \cY}
    \bigg( \sum_{x \in \cX}
    P_{XY}(x, y)^{\frac{1}{1 + \rho}}\bigg)^{1 + \rho},
\end{align*}
i.e.,
\begin{align*}
    E^{(3)}(\rho,P_{X})\geq E^{(1)}(\rho,P_{XY}).
\end{align*}
Hence, for any $n>0$ and $\rho \in [0, 1]$, we have
\begin{align}
    \max_{d\in\cD_n}E_n^{(1)}(\rho,P_{X^{n}Y_{(d)}^{n}})
    &= E^{(1)}(\rho,P_{XY}) + \frac{\max\{|\overline{d}_n|,|\underline{d}_n|\}}{n}\left( E^{(3)}(\rho,P_{X}) - E^{(1)}(\rho,P_{XY})\right)\notag\\
    &= E^{(1)}(\rho,P_{XY}) + \Delta_{\bfd}^{(n)}\left( E^{(3)}(\rho,P_{X}) - E^{(1)}(\rho,P_{XY})\right)\notag\\
    &= (1 - \Delta_{\bfd}^{(n)}) E^{(1)}(\rho,P_{XY}) + \Delta_{\bfd}^{(n)} E^{(3)}(\rho,P_{X}).
    \label{eq:ExBoundWithDelay1}
\end{align}
Then, we have
\begin{align}
    & \rho R_1-\max_{d\in\cD_n}E_n^{(1)}(\rho,P_{X^{n}Y_{(d)}^{n}})\notag\\
    &\overset{(a)}{=} \rho R_1  - (1-\Delta_{\bfd}^{(n)} )E^{(1)}(\rho,P_{XY})
    -\Delta_{\bfd}^{(n)}E^{(3)}(\rho,P_{X})\notag\\    
    &= \rho \left(R_1-(1-\Delta_{\bfd}^{(n)})H(P_{X|Y}^{(\rho)}|\bar P_{Y}^{(\rho)})-\Delta_{\bfd}^{(n)}H(P_{X}^{(\rho)}) +(1-\Delta_{\bfd}^{(n)})H(P_{X|Y}^{(\rho)}|\bar P_{Y}^{(\rho)}) + \Delta_{\bfd}^{(n)}H(P_{X}^{(\rho)})\right) \notag\\
    &\quad  - (1-\Delta_{\bfd}^{(n)} )E^{(1)}(\rho,P_{XY})
    -\Delta_{\bfd}^{(n)}E^{(3)}(\rho,P_{X})\notag\\
    &\overset{(b)}{=} \rho \left(R_1-(1-\Delta_{\bfd}^{(n)})H(P_{X|Y}^{(\rho)}|\bar P_{Y}^{(\rho)})-\Delta_{\bfd}^{(n)}H(P_{X}^{(\rho)})\right)\notag\\
    &\quad +(1-\Delta_{\bfd}^{(n)} )D(\bar P_{Y}^{(\rho)}\times P_{X|Y}^{(\rho)}\|P_{XY})  + \Delta_{\bfd}^{(n)}D(P_{X}^{(\rho)}\|P_{X}) \notag\\
    &= \rho \left(R_1 - H(P_{X|Y}^{(\rho)}|\bar P_{Y}^{(\rho)}) -\Delta_{\bfd}^{(n)}\left(H(P_{X}^{(\rho)}) - H(P_{X|Y}^{(\rho)}|\bar P_{Y}^{(\rho)})\right) \right)\notag\\
    &\quad + D(\bar P_{Y}^{(\rho)}\times P_{X|Y}^{(\rho)}\|P_{XY}) - \Delta_{\bfd}^{(n)} \left( D(\bar P_{Y}^{(\rho)}\times P_{X|Y}^{(\rho)}\|P_{XY}) - D(P_{X}^{(\rho)}\|P_{X})\right) \notag\\
    &\geq \rho \left(R_1 - H(P_{X|Y}^{(\rho)}|\bar P_{Y}^{(\rho)}) -\Delta_{\bfd}^{(n)} \left| H(P_{X}^{(\rho)}) - H(P_{X|Y}^{(\rho)}|\bar P_{Y}^{(\rho)})\right|_{+} \right)\notag\\
    &\quad + D(\bar P_{Y}^{(\rho)}\times P_{X|Y}^{(\rho)}\|P_{XY}) - \Delta_{\bfd}^{(n)} \left| D(\bar P_{Y}^{(\rho)}\times P_{X|Y}^{(\rho)}\|P_{XY}) - D(P_{X}^{(\rho)}\|P_{X})\right|_{+} \notag\\
    &= F_{1}(\rho, R_1, P_{XY}, \Delta_{\bfd}^{(n)}), \label{equ: ExBound1-2}
\end{align}
where (a) comes from (\ref{eq:ExBoundWithDelay1}) and (b) comes from Lemma \ref{lem: H - E = D}. By combining (\ref{equ: ExBound1-1}) and (\ref{equ: ExBound1-2}), we have (\ref{eq:ExBound1}) for the case where $i = 1$ as follows:
\begin{align*}
    \sup_{0\leq\rho\leq 1} \left(\rho R_1 - E_n^{(1)}(\rho,P_{\widetilde{X}^n\widetilde{Y}^n})\right)
    &\geq \sup_{0\leq\rho\leq 1} \min_{P_{XY}\in \bar{\cS}_{n}}
    \left(\rho R_1 - \max_{d\in \cD_n} E_n^{(1)}(\rho,P_{X^{n}Y_{(d)}^{n}})\right) - \epsilon'_n\notag\\
    &\geq \sup_{0\leq\rho\leq 1} \min_{P_{XY}\in \bar{\cS}_{n}} F_{1}(\rho, R_1, P_{XY}, \Delta_{\bfd}^{(n)}) - \epsilon'_n.
\end{align*}
Similarly, we have (\ref{eq:ExBound1}) for the case where $i = 2$.

Next, we show the bound (\ref{eq:ExBound1}) for the case where $i = 3$. We have
\begin{align}
    & E_n^{(3)}(\rho, P_{X^{n}Y_{(d)}^{n}})\notag\\
    &=
    \frac{1}{n}\log 
    \left( \sum_{x^n\in\cX^n} \sum_{y^n\in\cY^n}
        \left(P_{X}^{|d|}(x_{\rm I}^{|d|})
            P_{XY}^{n-|d|}(x_{\rm C}^{n-|d|}, y_{\rm C}^{n-|d|})
            P_{Y}^{|d|}(y_{\rm I}^{|d|})\right)^{\frac{1}{1+\rho}}\right)^{1+\rho}\notag\\
    &=
    \frac{1}{n}\log 
    \left( \sum_{x_{\rm I}^{|d|}\in\cX^{|d|}} 
        P_{X}^{|d|}(x_{\rm I}^{|d|})^{\frac{1}{1+\rho}}\right)^{1+\rho}
    \left(\sum_{x_{\rm C}^{n-|d|}\in\cX^{n-|d|}}\sum_{y_{\rm C}^{n-|d|}\in\cY^{n-|d|}}
        P_{XY}^{n-|d|}(x_{\rm C}^{n-|d|}, y_{\rm C}^{n-|d|})^{\frac{1}{1+\rho}}\right)^{1+\rho}\notag\\
    &\quad \times \left( \sum_{y_{\rm I}^{|d|}\in\cY^{|d|}}P_{Y}^{|d|}(y_{\rm I}^{|d|})^{\frac{1}{1+\rho}}\right)^{1+\rho}\notag\\
    &= \frac{|d|}{n}\log 
    \left( \sum_{x\in\cX} 
        P_{X}(x)^{\frac{1}{1+\rho}}\right)^{1+\rho}
    + \frac{n-|d|}{n}\log \left(\sum_{x\in\cX}\sum_{y\in\cY}
        P_{XY}(x, y)^{\frac{1}{1+\rho}}\right)^{1+\rho}
    +  \frac{|d|}{n}\log  \left( \sum_{y\in\cY}P_{Y}(y)^{\frac{1}{1+\rho}}\right)^{1+\rho}\notag\\
    &= \left(1 - \frac{|d|}{n}\right)\log \left(\sum_{x\in\cX}\sum_{y\in\cY}
        P_{XY}(x, y)^{\frac{1}{1+\rho}}\right)^{1+\rho} + \frac{|d|}{n} \left(\log 
        \left( \sum_{x\in\cX} 
            P_{X}(x)^{\frac{1}{1+\rho}}\right)^{1+\rho}
        +  \log  \left( \sum_{y\in\cY}P_{Y}(y)^{\frac{1}{1+\rho}}\right)^{1+\rho}\right)\notag\\
    &= \left(1 - \frac{|d|}{n}\right) E^{(3)}(\rho, P_{X Y}) + \frac{|d|}{n} \left(E^{(3)}(\rho, P_{X}) + E^{(3)}(\rho, P_{Y})\right).
    \label{eq:ExBoundWithDelay3}
\end{align}
On the other hand, for any $\rho \in [0, 1]$ and $d \in \cD_{n}$, we have,
\begin{align*}
    &H(P_{XY}^{(\rho)})+\Delta_{\bfd}^{(n)}\left|H(P_{X}^{(\rho)})+H(P_{Y}^{(\rho)}) - H(P_{XY}^{(\rho)})\right|_{+} \\
    & \geq  H(P_{XY}^{(\rho)}) +\frac{|d|}{n}\left|H(P_{X}^{(\rho)})+H(P_{Y}^{(\rho)}) - H(P_{XY}^{(\rho)})\right|_{+} \\
    &\geq \left(1-\frac{|d|}{n}\right)H(P_{XY}^{(\rho)}) +\frac{|d|}{n}\left(H(P_{X}^{(\rho)})+H(P_{Y}^{(\rho)})\right).
\end{align*}
Thus, we have
\begin{align}
    &\rho (R_1+R_2)-E_n^{(3)}(\rho,P_{X^{n}Y_{(d)}^{n}})\notag\\
    &= \rho \left(R_1+R_2 - \left(  H(P_{XY}^{(\rho)})+\Delta_{\bfd}^{(n)}\left|H(P_{X}^{(\rho)})+H(P_{Y}^{(\rho)}) - H(P_{XY}^{(\rho)})\right|_{+} \right) \right) \notag\\
    &\quad + \rho\left( H(P_{XY}^{(\rho)}) + \Delta_{\bfd}^{(n)}\left|H(P_{X}^{(\rho)}) + H(P_{Y}^{(\rho)}) -H(P_{XY}^{(\rho)})\right|_{+}\right) - E_n^{(3)}(\rho,P_{X^{n}Y_{(d)}^{n}})\notag\\
    &\geq \rho \left(R_1+R_2 - \left(  H(P_{XY}^{(\rho)}) + \Delta_{\bfd}^{(n)} \left|H(P_{X}^{(\rho)})+H(P_{Y}^{(\rho)}) - H(P_{XY}^{(\rho)})\right|_{+} \right) \right)\notag\\
    &\quad + \rho \left( \left( 1 - \frac{|d|}{n} \right) H(P_{XY}^{(\rho)}) + \frac{|d|}{n} \left(H(P_{X}^{(\rho)}) + H(P_{Y}^{(\rho)}) \right) \right) - E_n^{(3)}(\rho,P_{X^{n}Y_{(d)}^{n}}) \notag\\
    &\overset{(a)}{=} \rho \left(R_1+R_2 - \left(  H(P_{XY}^{(\rho)}) + \Delta_{\bfd}^{(n)} \left|H(P_{X}^{(\rho)})+H(P_{Y}^{(\rho)}) - H(P_{XY}^{(\rho)})\right|_{+} \right) \right)\notag\\
    &\quad + \rho \left( \left( 1 - \frac{|d|}{n} \right) H(P_{XY}^{(\rho)}) + \frac{|d|}{n} \left(H(P_{X}^{(\rho)}) + H(P_{Y}^{(\rho)}) \right) \right) \notag\\
    & \quad - \left(1 - \frac{|d|}{n}\right) E^{(3)}(\rho, P_{X Y}) - \frac{|d|}{n} \left(E^{(3)}(\rho, P_{X}) + E^{(3)}(\rho, P_{Y})\right)  \notag\\ 
    &\overset{\rm (b)}{=} \rho \left(R_1+R_2 - \left(  H(P_{XY}^{(\rho)}) + \Delta_{\bfd}^{(n)} \left|H(P_{X}^{(\rho)}) + H(P_{Y}^{(\rho)}) - H(P_{XY}^{(\rho)})\right|_{+} \right) \right)\notag\\
    &\quad + \left(1-\frac{|d|}{n}\right)D(P_{XY}^{(\rho)}\|P_{XY})
    + \frac{|d|}{n}\left(  D(P_{X}^{(\rho)}\|P_{X}) +  D(P_{Y}^{(\rho)}\|P_{Y}) \right) \notag\\
    & \geq \rho \left(R_1+R_2 - \left(  H(P_{XY}^{(\rho)}) + \Delta_{\bfd}^{(n)} \left|H(P_{X}^{(\rho)}) + H(P_{Y}^{(\rho)}) - H(P_{XY}^{(\rho)})\right|_{+} \right) \right)\notag\\ 
    &\quad +D(P_{XY}^{(\rho)}\|P_{XY})  - \frac{|d|}{n}\left| D(P_{XY}^{(\rho)}\|P_{XY})  - D(P_{X}^{(\rho)}\|P_{X}) - D(P_{Y}^{(\rho)}\|P_{Y}) \right|_{+} \notag\\
    & \geq \rho \left(R_1+R_2 - \left(  H(P_{XY}^{(\rho)}) + \Delta_{\bfd}^{(n)} \left|H(P_{X}^{(\rho)}) + H(P_{Y}^{(\rho)}) - H(P_{XY}^{(\rho)})\right|_{+} \right) \right)\notag\\
    &\quad + D(P_{XY}^{(\rho)}\|P_{XY})
    - \Delta_{\bfd}^{(n)}\left| D(P_{XY}^{(\rho)}\|P_{XY}) - D(P_{X}^{(\rho)}\|P_{X})  - D(P_{Y}^{(\rho)}\|P_{Y})  \right|_{+} \notag\\
    & = F_{3}(\rho,R_1 + R_2,P_{XY}, \Delta_{\bfd}^{(n)}),
    \label{inq:ExBound3-2}
\end{align}
where (a) comes from (\ref{eq:ExBoundWithDelay3}) and (b) comes from Lemma \ref{lem: H - E = D}. By combining (\ref{equ: ExBound1-1}) and (\ref{inq:ExBound3-2}),  we have (\ref{eq:ExBound1}) for the case where $i = 3$ as follows:
\begin{align*}
    \sup_{0\leq\rho\leq 1} \left(\rho (R_1+R_2) - E_n^{(3)}(\rho,P_{\widetilde{X}^n\widetilde{Y}^n})\right)
    &\geq \sup_{0\leq\rho\leq 1} \min_{P_{XY}\in \bar{\cS}_{n}}\min_{d\in \cD_n} \left(\rho (R_1+R_2)
        -E_n^{(3)}(\rho,P_{X^{n}Y_{(d)}^{n}})\right) - \epsilon'_n\\
    &\geq \sup_{0\leq\rho\leq 1} \min_{P_{XY}\in \bar{\cS}_{n}} \min_{d\in \cD_n}  F_{3}(\rho,R_1 + R_2,P_{XY}, \Delta_{\bfd}^{(n)}) -\epsilon'_n\\
    &= \sup_{0\leq\rho\leq 1} \min_{P_{XY}\in \bar{\cS}_{n}} F_{3}(\rho,R_1 + R_2,P_{XY}, \Delta_{\bfd}^{(n)}) -\epsilon'_n.
\end{align*}

\section{Proof of Theorem \ref{thm: positivity of E}}
\label{sec:proof of error exponent}
In this appendix, we prove Theorem \ref{thm: positivity of E}. To this end, we introduce some necessary lemmas.

\begin{lem} \label{lem: uniform continuity}
    For any $\rho \in [0, 1]$ and any $P_{XY} \in \cP(\cX \times \cY)$, we have
    \begin{align}
        d_{\rm v}(P_{X}, P_{X}^{(\rho)}) &\leq 2 \left(1 - |\cX|^{-\rho} \right), \label{lem: uniform continuity, eq1}\\
        d_{\rm v}(P_{XY}, \bar P_{Y}^{(\rho)} \times P_{X|Y}^{(\rho)}) &\leq 2 \left(1 - |\cX|^{-\rho} \right). \label{lem: uniform continuity, eq2}
    \end{align}
\end{lem}
This lemma shows that each variational distance uniformly converges to $0$ as $\rho \to 0$. The proof is given in Appendix \ref{appendix: uniform continuity}.

The next lemma follows immediately from \cite[Problem 3.10]{csiszar2011itc}.
\begin{lem}
    \label{lem: continuity of entropy}
    For any $\delta\in[0,1]$, and $P, Q\in\cP(\cX)$ satisfying $d_{\rm v}(P, Q) \leq \delta,$ we have
    \begin{align*}
        |H(P) - H(Q)| \leq \bar{\delta}(|\cX|,\delta),
    \end{align*}
    where $\bar{\delta}(N,\delta) = \frac{1}{2}\delta\log(N - 1) + h\left(\frac{\delta}{2}\right)$ if $N \geq 2$, $\bar{\delta}(1,\delta) = 0$, $h(p)= - (1-p)\log(1-p) - p\log p $, and note that $\bar{\delta}(N,\delta) \to 0$ as $\delta \to 0$. Furthermore, for any $\delta\in[0,1]$, $P, Q\in\cP(\cX)$, and $V,W\in\cP(\cY|\cX)$ satisfying $d_{\rm v}(P\times V, Q\times W) \leq \delta,$ we have
    \begin{align*}
        |H(V|P) - H(W|Q)| \leq 2\bar{\delta}(|\cX| |\cY|,\delta).
    \end{align*}
\end{lem}

Now, we prove Theorem \ref{thm: positivity of E}. Let $\gamma > 0$ be a constant such that
\begin{align}
    R_i &\geq \sup_{P_{XY}\in\cS} \left(H_{i}(X, Y)+\Delta I(X;Y)\right) + 3\gamma,\ \forall i \in \{1, 2, 3\}. \label{equ: cond 1 of error exponent}
\end{align}
We fix $\delta \in (0, 1]$ so that
\begin{align}
    (1 + \Delta ) 2 \bar{\delta}(|\cX| |\cY|,\delta) + \Delta  \bar{\delta}(|\cX|,\delta) &\leq \gamma, \label{equ: condition of delta 1}\\
    (1 + \Delta ) 2 \bar{\delta}(|\cX| |\cY|,\delta) + \Delta  \bar{\delta}(|\cY|,\delta) &\leq \gamma, \label{equ: condition of delta 2}\\
    \bar{\delta}(|\cX| |\cY|,\delta) + \Delta \left( \bar{\delta}(|\cX|,\delta) + \bar{\delta}(|\cY|,\delta) + \bar{\delta}(|\cX| |\cY|,\delta) \right) &\leq \gamma. \label{equ: condition of delta 3}
\end{align}
We also fix $\rho \in (0, 1]$ so that
\begin{align}
    2 \left(1 - |\cX|^{-\rho} \right) &\leq \delta, \label{equ: condition of rho 1}\\
    2 \left(1 - |\cY|^{-\rho} \right) &\leq \delta, \label{equ: condition of rho 2}\\
    2 \left(1 - (|\cX||\cY|)^{-\rho} \right) &\leq \delta. \label{equ: condition of rho 3}
\end{align}
We note that $\gamma > 0$ and $\delta, \rho \in (0, 1]$ are determined depending only on $(R_{1}, R_{2}, \cS, \Delta, |\cX|, |\cY|)$. For these constants $\gamma, \delta$ and $\rho$, we will show that
\begin{align}
    R_1 &\geq \sup_{P_{XY}\in\cS_{\delta}} \left(H(P_{X|Y}^{(\rho)}|\bar P_{Y}^{(\rho)}) + \Delta \left| H(P_{X}^{(\rho)}) - H(P_{X|Y}^{(\rho)}|\bar P_{Y}^{(\rho)}) \right|_{+}\right) + \gamma, \label{equ: R1 geq H with rho^*}\\
    R_2 &\geq \sup_{P_{XY}\in\cS_{\delta}} \left(H(P_{Y|X}^{(\rho)}|\bar P_{X}^{(\rho)}) + \Delta \left| H(P_{Y}^{(\rho)}) - H(P_{Y|X}^{(\rho)}|\bar P_{X}^{(\rho)}) \right|_{+}\right) + \gamma, \label{equ: R2 geq H with rho^*}\\
    R_1 + R_2 &\geq \sup_{P_{XY}\in\cS_{\delta}} \left(H(P_{XY}^{(\rho)}) + \Delta  \left|H(P_{X}^{(\rho)}) + H(P_{Y}^{(\rho)}) - H(P_{XY}^{(\rho)})\right|_{+}\right) + \gamma. \label{equ: R1 + R2 geq H with rho^*}
\end{align}
Then, by using these inequalities, we can easily show that $F_{i}(\rho, R_i, P_{XY},\Delta) \geq \rho \gamma > 0$ for any $P_{XY} \in \cS_{\delta}$ and $i \in \{1, 2, 3\}$. This implies \eqref{thm: equ: positivity of F}. Here, we use the fact that for \eqref{eq: ExFunction1}--\eqref{eq: ExFunction3},
\begin{align*}
    D(\bar P_{Y}^{(\rho)}\times P_{X|Y}^{(\rho)}\|P_{XY}) - \Delta \left|D(\bar P_{Y}^{(\rho)}\times P_{X|Y}^{(\rho)}\|P_{XY}) - D(P_{X}^{(\rho)}\|P_{X}) \right|_{+} &\geq 0,\\
    D(\bar P_{X}^{(\rho)}\times P_{Y|X}^{(\rho)}\|P_{XY}) - \Delta\left| D(\bar P_{X}^{(\rho)}\times P_{Y|X}^{(\rho)}\|P_{XY}) - D(P_{Y}^{(\rho)}\|P_{Y})\right|_{+} &\geq 0,\\
    D(P_{XY}^{(\rho)}\|P_{XY}) - \Delta\left| D(P_{XY}^{(\rho)}\|P_{XY}) - D(P_{X}^{(\rho)}\|P_{X}) - D(P_{Y}^{(\rho)}\|P_{Y}) \right|_{+} &\geq 0.
\end{align*}
Hence, all we need is to show \eqref{equ: R1 geq H with rho^*}--\eqref{equ: R1 + R2 geq H with rho^*}.

From the definition of $\cS_{\delta}$, for any $P_{XY}\in\cS_{\delta}$, there exists $P_{X'Y'}\in\cS$ such that $d_{\rm v}(P_{XY},P_{X'Y'})\leq \delta$. Hence, according to Lemma \ref{lem: continuity of entropy}, we have for any $i \in \{1, 2, 3\}$,
\begin{align*}
    H_{i}(X', Y') + \Delta I(X';Y')
    & \geq  H_{i}(X, Y) + \Delta I(X;Y) - \left( (1 + \Delta ) 2 \bar{\delta}(|\cX| |\cY|,\delta) + \Delta  \bar{\delta}(|\cX|,\delta) \right).
\end{align*}
By recalling that $\delta\in (0,1]$ satisfies \eqref{equ: condition of delta 1}, we have for any $i \in \{1, 2, 3\}$,
\begin{align}
    \sup_{P_{XY}\in\cS} \left(H_{i}(X, Y) + \Delta I(X;Y)\right) \geq \sup_{P_{XY}\in\cS_{\delta}} \left(H_{i}(X, Y) + \Delta I(X;Y)\right) - \gamma. \label{equ: positivity of error exponent step 1-1}
\end{align}

On the other hand, according to Lemma \ref{lem: uniform continuity} and \eqref{equ: condition of rho 1}--\eqref{equ: condition of rho 3}, for any $P_{XY} \in \cP(\cX \times \cY)$, we have
\begin{align*}
    d_{\rm v}(P_{X}, P_{X}^{(\rho)}) &\leq 2 \left(1 - |\cX|^{-\rho} \right) \leq \delta,\\
    d_{\rm v}(P_{Y}, P_{Y}^{(\rho)}) &\leq 2 \left(1 - |\cY|^{-\rho} \right) \leq \delta,\\
    d_{\rm v}(P_{XY}, P_{XY}^{(\rho)}) &\leq 2 \left(1 - (|\cX||\cY|)^{-\rho} \right) \leq \delta,\\
    d_{\rm v}(P_{XY}, \bar P_{Y}^{(\rho)} \times P_{X|Y}^{(\rho)}) &\leq 2 \left(1 - |\cX|^{-\rho} \right) \leq \delta,\\
    d_{\rm v}(P_{XY}, \bar P_{X}^{(\rho)} \times P_{Y|X}^{(\rho)}) &\leq 2 \left(1 - |\cY|^{-\rho} \right) \leq \delta.
\end{align*}
Hence, according to Lemma \ref{lem: continuity of entropy}, we have
\begin{align*}
    H(X|Y) + \Delta I(X; Y)
    &= H(P_{X|Y}|P_{Y}) + \Delta \left|H(P_{X}) - H(P_{X|Y}|P_{Y})\right|_{+}\\
    &\geq H(P_{X|Y}^{(\rho)}|\bar P_{Y}^{(\rho)}) + \Delta \left|H(P_{X}^{(\rho)}) - H(P_{X|Y}^{(\rho)}|\bar P_{Y}^{(\rho)}) \right|_{+}\\
    &\quad  - \left( (1 + \Delta ) 2 \bar{\delta}(|\cX| |\cY|,\delta) + \Delta  \bar{\delta}(|\cX|,\delta) \right),\\
    H(Y|X) + \Delta I(X; Y)
    &\geq H(P_{Y|X}^{(\rho)}|\bar P_{X}^{(\rho)}) + \Delta \left|H(P_{Y}^{(\rho)}) - H(P_{Y|X}^{(\rho)}|\bar P_{X}^{(\rho)}) \right|_{+}\\
    &\quad  - \left( (1 + \Delta ) 2 \bar{\delta}(|\cX| |\cY|,\delta) + \Delta  \bar{\delta}(|\cY|,\delta) \right),\\
    H(X, Y) + \Delta I(X; Y)
    &=  H(P_{XY}) + \Delta \left| H(P_{X}) + H(P_{Y}) - H(P_{XY}) \right|_{+}\\
    & \geq H(P_{XY}^{(\rho)}) + \Delta  \left|H(P_{X}^{(\rho)}) + H(P_{Y}^{(\rho)}) - H(P_{XY}^{(\rho)})\right|_{+} \\
    &\quad - \left( \bar{\delta}(|\cX| |\cY|,\delta) + \Delta \left( \bar{\delta}(|\cX|,\delta) + \bar{\delta}(|\cY|,\delta) + \bar{\delta}(|\cX| |\cY|,\delta) \right) \right).
\end{align*}
By recalling that $\delta > 0$ satisfies \eqref{equ: condition of delta 1}--\eqref{equ: condition of delta 3}, we have
\begin{align}
    \sup_{P_{XY}\in\cS_{\delta}} \left(H(X|Y) + \Delta I(X;Y)\right)
    &\geq \sup_{P_{XY} \in \cS_{\delta}} \left(H(P_{X|Y}^{(\rho)}|\bar P_{Y}^{(\rho)}) + \Delta \left| H(P_{X}^{(\rho)}) - H(P_{X|Y}^{(\rho)}|\bar P_{Y}^{(\rho)}) \right|_{+}\right) - \gamma,\label{equ: positivity of error exponent step 2-1}\\
    \sup_{P_{XY}\in\cS_{\delta}} \left(H(Y|X) + \Delta I(X;Y)\right)
    &\geq \sup_{P_{XY} \in \cS_{\delta}} \left(H(P_{Y|X}^{(\rho)}|\bar P_{X}^{(\rho)}) + \Delta \left| H(P_{Y}^{(\rho)}) - H(P_{Y|X}^{(\rho)}|\bar P_{X}^{(\rho)}) \right|_{+}\right) - \gamma, \label{equ: positivity of error exponent step 2-2}\\
    \sup_{P_{XY}\in\cS_{\delta}} \left(H(X,Y) + \Delta I(X;Y)\right)
    &\geq \sup_{P_{XY} \in \cS_{\delta}} \left(H(P_{XY}^{(\rho)}) + \Delta  \left|H(P_{X}^{(\rho)}) + H(P_{Y}^{(\rho)}) - H(P_{XY}^{(\rho)})\right|_{+}\right) - \gamma. \label{equ: positivity of error exponent step 2-3}
\end{align}

As a consequence of inequalities \eqref{equ: positivity of error exponent step 1-1}--\eqref{equ: positivity of error exponent step 2-3}, we have
\begin{align*}
    \sup_{P_{XY}\in\cS} \left(H(X|Y) + \Delta I(X;Y)\right)
    &\geq \sup_{P_{XY}\in\cS_{\delta}} \left(H(P_{X|Y}^{(\rho)}|\bar P_{Y}^{(\rho)}) + \Delta \left|H(P_{X}^{(\rho)}) - H(P_{X|Y}^{(\rho)}|\bar P_{Y}^{(\rho)}) \right|_{+}\right) - 2\gamma,\\
    \sup_{P_{XY}\in\cS} \left(H(Y | X) + \Delta I(X;Y)\right)
    &\geq \sup_{P_{XY}\in\cS_{\delta}} \left(H(P_{Y|X}^{(\rho)}|\bar P_{X}^{(\rho)}) + \Delta \left|H(P_{Y}^{(\rho)}) - H(P_{Y|X}^{(\rho)}|\bar P_{X}^{(\rho)}) \right|_{+}\right) - 2\gamma,\\
    \sup_{P_{XY}\in\cS} \left(H(X, Y) + \Delta I(X;Y)\right)
    &\geq \sup_{P_{XY}\in\cS_{\delta}} \left(H(P_{XY}^{(\rho)}) + \Delta  \left|H(P_{X}^{(\rho)}) + H(P_{Y}^{(\rho)}) - H(P_{XY}^{(\rho)})\right|_{+}\right) - 2\gamma.
\end{align*}
Hence, due to these inequalities and \eqref{equ: cond 1 of error exponent}, we have \eqref{equ: R1 geq H with rho^*}--\eqref{equ: R1 + R2 geq H with rho^*}.

\section{Proof of Lemma \ref{lem: uniform continuity}}
\label{appendix: uniform continuity}
In order to prove Lemma \ref{lem: uniform continuity}, we give some inequalities in the following lemmas and corollary.

\begin{lem}[{\cite[Theorem 27]{hardy1952inequalities}}]
    \label{lem: hardy's inequality}
    For any $x,y\geq 0$, and $r \in [0, 1]$, we have
    \begin{align*}
        (x + y)^r \leq x^r + y^r.
    \end{align*}
\end{lem}

\begin{lem}
    \label{lem: bounds for Renyi type prob}
    For any $P_{X} \in \cP(\cX)$ and any $\rho \in [0, 1]$, we have
    \begin{align*}
        1 \leq \left(\sum_{x\in\cX} P_{X}(x)^{\frac{1}{1 + \rho}}\right)^{1 + \rho} \leq |\cX|^{\rho}.
    \end{align*} 
\end{lem}
\begin{IEEEproof}
    The first inequality comes from Lemma \ref{lem: hardy's inequality} and the monotonicity of the function $f(a) = a^{1 + \rho}$ for $a \geq 0$. The second inequality comes from Jensen's inequality as follows:
    \begin{flalign*}
        & & \left(\sum_{x\in\cX} P_{X}(x)^{\frac{1}{1 + \rho}}\right)^{1 + \rho}
        &= |\cX|^{1 + \rho} \left(\sum_{x\in\cX} \frac{1}{|\cX|} P_{X}(x)^{\frac{1}{1 + \rho}}\right)^{1 + \rho}\\
        & & &\leq |\cX|^{1 + \rho} \sum_{x\in\cX} \frac{1}{|\cX|} P_{X}(x)
        = |\cX|^{\rho}. & \IEEEQEDhere
    \end{flalign*}
\end{IEEEproof}
\begin{cor}
    \label{cor: bounds for Renyi type prob}
    For any $P_{XY} \in \cP(\cX \times \cY)$ and any $\rho \in [0, 1]$, we have
    \begin{align*}
        P_{Y}(y) \leq \left(\sum_{x\in\cX} P_{XY}(x,y)^{\frac{1}{1+\rho}}\right)^{1+\rho} \leq P_{Y}(y) |\cX|^{\rho},\ \forall y \in \cY,
    \end{align*}
    and hence
    \begin{align*}
        1 \leq \sum_{y\in\cY}\left(\sum_{x\in\cX} P_{XY}(x,y)^{\frac{1}{1+\rho}}\right)^{1+\rho} \leq |\cX|^{\rho}.
    \end{align*}
\end{cor}

Now, we show \eqref{lem: uniform continuity, eq1} and \eqref{lem: uniform continuity, eq2}. First, we show \eqref{lem: uniform continuity, eq1}. For any $\rho \in [0, 1]$ and any $x \in \cX$, we have
\begin{align}
    P_{X}^{(\rho)}(x) &= \frac{P_{X}(x)^{\frac{1}{1 + \rho}}}{\sum_{x\in\cX}P_{X}(x)^{\frac{1}{1 + \rho}}} \notag \\
    &\geqo{(a)} \frac{P_{X}(x)}{\left(\sum_{x\in\cX}P_{X}(x)^{\frac{1}{1 + \rho}}\right)^{1 + \rho}} \notag \\
    &\geqo{(b)} P_{X}(x) |\cX|^{-\rho}, \label{equ:bound_of_Pxrho}
\end{align}
where (a) comes from elementary inequalities $P_{X}(x)^{\frac{1}{1 + \rho}} \geq P_{X}(x)$ and $\sum_{x\in\cX}P_{X}(x)^{\frac{1}{1 + \rho}} \leq \left(\sum_{x\in\cX}P_{X}(x)^{\frac{1}{1 + \rho}}\right)^{1 + \rho}$, and (b) comes from Lemma \ref{lem: bounds for Renyi type prob}. Thus, we have \eqref{lem: uniform continuity, eq1} as follows:
\begin{align*}
    d_{\rm v}(P_{X}, P_{X}^{(\rho)}) &= \sum_{x\in\cX} |P_{X}(x) - P_{X}^{(\rho)}(x)|\\
    &= 2 \sum_{x\in\cX: P_{X}(x) \geq P_{X}^{(\rho)}(x)} \left( P_{X}(x) - P_{X}^{(\rho)}(x) \right)\\
    &\leqo{(a)} 2 \sum_{x\in\cX: P_{X}(x) \geq P_{X}^{(\rho)}(x)} \left( P_{X}(x) - P_{X}(x) |\cX|^{-\rho} \right)\\
    &= 2 \left( 1 - |\cX|^{-\rho} \right) \sum_{x\in\cX: P_{X}(x) \geq P_{X}^{(\rho)}(x)} P_{X}(x)\\
    &\leq 2 \left( 1 - |\cX|^{-\rho} \right),
\end{align*}
where (a) comes from \eqref{equ:bound_of_Pxrho}.

Next, we show \eqref{lem: uniform continuity, eq2}. For any $\rho \in [0, 1]$ and any $(x, y) \in \cX \times \cY$, we have
\begin{align}
    \bar P_{Y}^{(\rho)}(y) P_{X|Y}^{(\rho)}(x|y)
    &= \frac{\left(\sum_{x\in\cX} P_{XY}(x,y)^{\frac{1}{1 + \rho}}\right)^{1 + \rho}}{\sum_{y\in\cY}\left(\sum_{x\in\cX} P_{XY}(x,y)^{\frac{1}{1 + \rho}}\right)^{1 + \rho}}
    \frac{P_{XY}(x,y)^{\frac{1}{1 + \rho}}} {\sum_{x\in\cX} P_{XY}(x,y)^{\frac{1}{1 + \rho}}}\notag\\
    &= \frac{P_{Y}(y) \left(\sum_{x\in\cX} P_{X|Y}(x|y)^{\frac{1}{1 + \rho}}\right)^{1 + \rho}}{\sum_{y\in\cY}\left(\sum_{x\in\cX} P_{XY}(x,y)^{\frac{1}{1 + \rho}}\right)^{1 + \rho}}
    \frac{P_{Y}(y)^{\frac{1}{1 + \rho}} P_{X|Y}(x|y)^{\frac{1}{1 + \rho}}}{P_{Y}(y)^{\frac{1}{1 + \rho}} \sum_{x\in\cX} P_{X|Y}(x|y)^{\frac{1}{1 + \rho}}}\notag\\
    &= \frac{P_{Y}(y) \left(\sum_{x\in\cX} P_{X|Y}(x|y)^{\frac{1}{1 + \rho}}\right)^{1 + \rho}}{\sum_{y\in\cY}\left(\sum_{x\in\cX} P_{XY}(x,y)^{\frac{1}{1 + \rho}}\right)^{1 + \rho}}
    \frac{P_{X|Y}(x|y)^{\frac{1}{1 + \rho}}}{\sum_{x\in\cX} P_{X|Y}(x|y)^{\frac{1}{1 + \rho}}}\notag\\
    &\geqo{(a)} \frac{P_{Y}(y) \left(\sum_{x\in\cX} P_{X|Y}(x|y)^{\frac{1}{1 + \rho}}\right)^{1 + \rho}}{\sum_{y\in\cY}\left(\sum_{x\in\cX} P_{XY}(x,y)^{\frac{1}{1 + \rho}}\right)^{1 + \rho}}
    \frac{P_{X|Y}(x|y)}{\left( \sum_{x\in\cX} P_{X|Y}(x|y)^{\frac{1}{1 + \rho}} \right)^{1 + \rho}}\notag\\
    &= \frac{P_{XY}(x, y)}{\sum_{y\in\cY}\left(\sum_{x\in\cX} P_{XY}(x,y)^{\frac{1}{1 + \rho}}\right)^{1 + \rho}}\notag\\
    &\geqo{(b)} P_{XY}(x, y) |\cX|^{-\rho},
    \label{equ:bound_of_PyrhoPxyrho}
\end{align}
where (a) again comes from the same type of elementary inequalities as before, and (b) comes from Corollary \ref{cor: bounds for Renyi type prob}. Thus, we have \eqref{lem: uniform continuity, eq2} as follows:
\begin{align*}
    d_{\rm v}(P_{XY}, \bar P_{Y}^{(\rho)} \times P_{X|Y}^{(\rho)})
    &= \sum_{(x, y) \in \cX \times \cY} |P_{XY}(x, y) - \bar P_{Y}^{(\rho)}(y) P_{X|Y}^{(\rho)}(x|y)|\\
    &= 2 \sum_{ \substack{(x, y) \in \cX \times \cY:\\ P_{XY}(x, y) \geq \bar P_{Y}^{(\rho)}(y) P_{X|Y}^{(\rho)}(x|y)} } \left( P_{XY}(x, y) - \bar P_{Y}^{(\rho)}(y) P_{X|Y}^{(\rho)}(x|y)\right)\\
    &\leqo{(a)} 2 \sum_{ \substack{(x, y) \in \cX \times \cY:\\ P_{XY}(x, y) \geq \bar P_{Y}^{(\rho)}(y) P_{X|Y}^{(\rho)}(x|y)} } \left( P_{XY}(x, y) - P_{XY}(x, y) |\cX|^{-\rho}\right)\\
    &= 2 \left(1 - |\cX|^{-\rho} \right) \sum_{ \substack{(x, y) \in \cX \times \cY:\\ P_{XY}(x, y) \geq \bar P_{Y}^{(\rho)}(y) P_{X|Y}^{(\rho)}(x|y)} } P_{XY}(x, y)\\
    &\leq 2 \left(1 - |\cX|^{-\rho} \right),
\end{align*}
where (a) comes from \eqref{equ:bound_of_PyrhoPxyrho}.


\ifCLASSOPTIONcaptionsoff
\newpage
\fi

\end{document}